\documentclass[12pt,preprint]{aastex}
\usepackage{natbib}

\def\mas{{\rm mas}}
\def\masyr{{\rm mas}\,{\rm yr}^{-1}}
\def\kms{{\rm km}\,{\rm s}^{-1}}
\def\au{{\rm au}} 
\def\e{{\rm E}}
\def\rel{{\rm rel}}
\def\bpi{{\bdv\pi}}
\def\bmu{{\bdv\mu}}
\def\bv{{\bdv v}}
\def\hel{{\rm hel}}
\def\geo{{\rm geo}}
\def\lsr{{\rm lsr}}
\def\rot{{\rm rot}}
\def\kpc{{\rm kpc}}

\newcommand{\bdv}[1]{\mbox{\boldmath$#1$}}

\begin{document}

\title{\emph{Spitzer} Microlensing Parallax for OGLE-2016-BLG-1067:\\
a sub-Jupiter Orbiting an M-dwarf in the Disk}

\author
{
S.~Calchi~Novati$^{1}$,
D.~Suzuki$^{2,3}$, 
A.~Udalski$^{4}$, 
A.~Gould$^{5,6,7}$,
Y.~Shvartzvald$^{8,*}$,
V.~Bozza$^{9,10}$,
D.~P.~Bennett$^{2}$,\\
and\\
C.~Beichman$^{11}$, 
G.~Bryden$^{8}$, 
S.~Carey$^{12}$, 
B.~S.~Gaudi$^{6}$, 
C.~B.~Henderson$^{11}$, 
J.~C.~Yee$^{13}$, 
W.~Zhu$^{6}$\\
(\emph{Spitzer} team)\\
and\\
F.~Abe$^{14}$, 
Y.~Asakura$^{14}$, 
R.~Barry$^{2}$,
A.~Bhattacharya$^{2,15}$, 
I.~A.~Bond$^{16}$,
M.~Donachie$^{17}$,
P.~Evans$^{17}$, 
A.~Fukui$^{18}$, 
Y.~Hirao$^{19}$, 
Y.~Itow$^{14}$,
K.~Kawasaki$^{19}$, 
N.~Koshimoto$^{19}$, 
M.~C.~A.~Li$^{17}$,
C.~H.~Ling$^{16}$, 
Y.~Matsubara$^{14}$,
S.~Miyazaki$^{19}$, 
Y.~Muraki$^{14}$, 
M.~Nagakane$^{19}$,
K.~Ohnishi$^{20}$, 
C.~Ranc$^{2,*}$, 
N.~J.~Rattenbury$^{17}$,
To.~Saito$^{21}$, 
A.~Sharan$^{17}$, 
D.~J.~Sullivan$^{22}$,
T.~Sumi$^{19}$,
P.~J.~Tristram$^{23}$,
T.~Yamada$^{19}$, 
A.~Yonehara$^{24}$\\
(MOA Collaboration)\\
and\\
P.~Mr\'{o}z$^{4}$, 
R.~Poleski$^{4,6}$, 
J.~Skowron$^{4}$, 
M.~K.~Szyma\'{n}ski$^{4}$, 
I.~Soszy\'{n}ski$^{4}$, 
S.~Koz{\l}owski$^{4}$,
P.~Pietrukowicz$^{4}$, 
K.~Ulaczyk$^{4}$,
M.~Pawlak$^{4}$\\
(OGLE Collaboration)\\
and\\
M.~D.~Albrow$^{25}$,
S.-J.~Chung$^{5,26}$, 
C.~Han$^{27}$, 
K.-H.~Hwang$^{5}$,
Y.~K.~Jung$^{13}$, 
Y.-H.~Ryu$^{5}$,
I.-G.~Shin$^{13}$,
W.~Zang$^{28,29}$,
S.-M.~Cha$^{5,30}$, 
D.-J.~Kim$^{5}$, 
H.-W.~Kim$^{5}$, 
S.-L.~Kim$^{5,26}$, 
C.-U.~Lee$^{5,26}$,
D.-J.~Lee$^{5}$,
Y.~Lee$^{5,30}$, 
B.-G.~Park$^{5,26}$,
R.~W.~Pogge$^{6}$ \\
(KMTNet Collaboration)\\
%
\normalsize{$^{1}$ IPAC, Mail Code 100-22,
Caltech, 1200 E. California Blvd., Pasadena, CA 91125, USA}\\
\normalsize{$^{2}$ Code 667, NASA Goddard Space Flight Center, Greenbelt,
MD 20771, USA}\\
\normalsize{$^{3}$ Institute of Space and Astronautical Science, Japan
Aerospace Exploration Agency, Kanagawa 252-5210, Japan}\\
\normalsize{$^{4}$ Warsaw University Observatory, Al.~Ujazdowskie~4, 00-478~Warszawa, Poland} \\ 
\normalsize{$^{5}$ Korea Astronomy and Space Science Institute, Daejon
34055, Korea}\\
\normalsize{$^{6}$ Department of Astronomy, Ohio State University, 140 W.
18th Ave., Columbus, OH 43210, USA}\\
\normalsize{$^{7}$ Max-Planck-Institute for Astronomy, K\"{o}nigstuhl 17,
69117 Heidelberg, Germany}\\
\normalsize{$^{8}$ Jet Propulsion Laboratory, California Institute of
Technology, 4800 Oak Grove Drive, Pasadena, CA 91109, USA} \\
\normalsize{$^{9}$ Dipartimento di Fisica ``E. R. Caianiello'', Universit\`a di Salerno, Via Giovanni Paolo II, 84084 Fisciano (SA),\ Italy} \\
\normalsize{$^{10}$ Istituto Nazionale di Fisica Nucleare, Sezione di Napoli, Via Cintia, 80126 Napoli, Italy}\\
\normalsize{$^{11}$ NASA Exoplanet Science Institute, California Institute
of Technology, Pasadena, CA 91125, USA}\\
\normalsize{$^{12}$ Spitzer, Science Center, MS 220-6, California
Institute of Technology,Pasadena, CA, USA}\\
\normalsize{$^{13}$ Harvard-Smithsonian Center for Astrophysics,
60 Garden St., Cambridge, MA 02138, USA}\\
\normalsize{$^{14}$ Institute for Space-Earth Environmental Research,
Nagoya University, Nagoya 464-8601, Japan}\\
\normalsize{$^{15}$ Department of Astronomy, University of Maryland, College Park, Maryland, USA}\\
\normalsize{$^{16}$ Institute of Natural and Mathematical Sciences, Massey
University, Auckland 0745, New Zealand}\\
\normalsize{$^{17}$ Department of Physics, University of Auckland, Private
Bag 92019, Auckland, New Zealand}\\
\normalsize{$^{18}$ Okayama Astrophysical Observatory, National
Astronomical Observatory of Japan, 3037-5 Honjo, Kamogata, Asakuchi,
Okayama 719-0232, Japan}\\
\normalsize{$^{19}$ Department of Earth and Space Science, Graduate School of Science, Osaka University, Toyonaka, Osaka 560-0043, Japan}\\
\normalsize{$^{20}$ Nagano National College of Technology, Nagano
381-8550, Japan}\\
\normalsize{$^{21}$ Tokyo Metropolitan College of Aeronautics, Tokyo
116-8523, Japan}\\
\normalsize{$^{22}$ School of Chemical and Physical Sciences, Victoria
University, Wellington, New Zealand}\\
\normalsize{$^{23}$ University of Canterbury Mt.\ John Observatory, P.O.
Box 56, Lake Tekapo 8770, New Zealand}\\
\normalsize{$^{24}$ Department of Physics, Faculty of Science, Kyoto
Sangyo University, 603-8555 Kyoto, Japan}\\
\normalsize{$^{25}$ University of Canterbury, Department of Physics and
Astronomy, Private Bag 4800, Christchurch 8020, New Zealand}\\
\normalsize{$^{26}$ Korea  University of Science and Technology, 
217 Gajeong-ro, Yuseong-gu, Daejeon 34113, Korea}\\
\normalsize{$^{27}$ Department of Physics, Chungbuk National University,
Cheongju 28644, Republic of Korea}\\
\normalsize{$^{28}$ Physics Department and Tsinghua Centre for
Astrophysics, Tsinghua University, Beijing 100084, China}\\
\normalsize{$^{29}$ Department of Physics, Zhejiang University, Hangzhou,
310058, China}\\
\normalsize{$^{30}$ School of Space Research, Kyung Hee University,
Yongin, Kyeonggi 17104, Korea}\\
}
\altaffiltext{*}{NASA Postdoctoral Program Fellow}


\newpage

\begin{abstract}
We report the discovery of a sub-Jupiter mass planet
orbiting beyond the snow line of an M-dwarf 
most likely in the Galactic disk  as part of the joint 
\emph{Spitzer} and ground-based monitoring
of microlensing planetary anomalies toward the Galactic bulge.
The microlensing parameters are strongly constrained
by the light curve modeling and in particular
by the \emph{Spitzer}-based measurement of the microlens
parallax, $\pi_\mathrm{E}$.  However,
in contrast to many planetary microlensing events, there are
no caustic crossings, so the angular
Einstein radius, $\theta_\mathrm{E}$ 
has only an upper limit based on the light curve modeling alone. 
Additionally, the analysis leads us to identify 8 degenerate configurations:
the four-fold microlensing parallax degeneracy being doubled
by a degeneracy in the caustic structure present at the level
of the ground-based solutions. 
To pinpoint the physical parameters, and at the same time
to break  the parallax degeneracy,
we make use of a series of arguments:
the $\chi^2$ hierarchy, the Rich argument,
and a prior Galactic model.
The preferred configuration is for a host 
at $D_L=3.73_{-0.67}^{+0.66}~\mathrm{kpc}$ with mass
$M_\mathrm{L}=0.30_{-0.12}^{+0.15}~\mathrm{M_\odot}$, orbited by a Saturn-like
planet with $M_\mathrm{planet}=0.43_{-0.17}^{+0.21}~\mathrm{M_\mathrm{Jup}}$
at projected separation $a_\perp = 1.70_{-0.39}^{+0.38}~\mathrm{au}$,
about 2.1 times beyond the system snow line.
Therefore, it adds to the growing
population of sub-Jupiter planets orbiting
near or beyond the snow line of M-dwarfs discovered by microlensing.
Based on the rules of the real-time protocol for the selection of events
to be followed up with \emph{Spitzer},
this planet will \emph{not} enter the sample for measuring
the Galactic distribution of planets.
\end{abstract}



\section{Introduction}
\label{sec:intro}


The {\it Spitzer} satellite is conducting a 5 year campaign (2014-18)
to measure the ``microlens parallax'' of about 750 microlensing
events toward the Galactic bulge by taking advantage of {\it Spitzer}'s
roughly 1~au projected separation from Earth
\citep{spitzer13,spitzer14,spitzer15a,spitzer15b,spitzer16}.  
The main goal of this program is to measure or constrain the mass and distance
of lens systems that contain planets.  This is why the target 
selection is designed to maximize event sensitivity to planets \citep{yee15},
with 4 planetary systems already characterized 
\citep{ob140124,ob150966,ob161195,ob161190}.
At the same time the survey is also probing 
a wide variety of other key science questions,
including massive remnants \citep{ob151285}, binary brown dwarfs
\citep{ob161469}, and the low-mass isolated-object mass function
\citep{zhu16,ob151482}.

The microlens parallax $\bpi_\e$ is a vector that quantifies the 
displacement of the lens-source separation in the Einstein ring
due to a displacement of the observer,
\begin{equation} \label{eq:pie}
\bpi_\e \equiv \pi_\e\,{\bmu\over \mu};
\qquad \pi_\e\equiv {\pi_\rel\over\theta_\e} = \sqrt{\pi_\rel\over \kappa M};
\qquad \kappa\equiv {4G\over c^2\au}\simeq 8.144\,{\textrm{mas}\over M_\odot},
\end{equation}
where $M$ is the lens mass, and
$\bmu$ and $\pi_\rel$ are respectively the lens-source relative
proper motion and parallax (we refer to \citealt{gould00b}
for an introduction to the formalism of microlensing).

For a substantial majority of published microlensing planets, $\theta_\e$
is measured because the planet is only noticed by the passage of the
source close to a caustic. If the planet actually transits the 
caustic (or comes very close), then it is possible to measure the
source radius crossing time, $t_*$, which is related to the Einstein
radius by 
\begin{equation} \label{eq:mu}
\mu = {\theta_*\over t_*} = {\theta_\e\over t_\e},
\end{equation}
where $\theta_*$ is the source angular radius, $\theta_\e$ 
the Einstein angular radius, and $t_\e$ 
is the Einstein timescale (which is well measured for almost all events).
For this subclass of events, the addition of a parallax measurement
directly yields
\begin{equation} \label{eq:mlens}
M = {\theta_\e\over\kappa\pi_\e}\,;
\qquad
\pi_\rel = \theta_\e\pi_\e\,.  
\end{equation}
However, \cite{zhu14} argued that in the era of pure-survey detection
of microlensing planets, only about half of the planets that are robustly
detected would yield $\theta_\e$ measurements.  Hence, there is a real
question of what can be said about the mass and distance of planets with
{\it Spitzer} parallax measurements if $\theta_\e$ is unknown.

In fact, there is a substantial amount of work that bears on this
question, mostly related to point-lens events, for which $\theta_\e$
measurements are extremely rare. \cite{hangould95} argued that while
$\pi_\e$ and $\theta_\e$ appear symmetrically in Equation~(\ref{eq:mlens}), 
the parallax information is intrinsically more valuable.  
This is basically because
the great majority of microlenses have proper motions spanning a range of
a factor 3, $2\,{\rm mas\, yr^{-1}}<\mu<6\,{\rm mas\, yr^{-1}}$.  Hence
if one simply guesses $\mu=4\,{\rm mas\, yr^{-1}}$, one already has a 
pretty good estimate of $\theta_\e=\mu t_\e$.  Therefore, actually measuring
$\theta_\e$ adds relatively little statistical information, although
it can be extremely important in the handful of cases that $\mu$ lies
substantially outside this range\footnote{Here we refer specifically
to photometric microlensing: in the future, astrometric microlensing, 
e.g., \cite{gouldyee14},
or interferometric observation of microlensing events, \cite{cassan16},
may provide crucial independent measurements of $\theta_\e$.}.  
By the same token, this means that
a measurement of $\pi_\e$ by itself can give a good estimate of the mass:
$M\sim (4\,{\rm mas\,yr^{-1}})t_\e/\kappa\pi_\e$. \cite{hangould95} did not
restrict themselves to such qualitative arguments but showed, using
their eponymous Galactic model, that distances could be quite well
constrained.

\cite{21event} and \cite{zhu17a} applied variants of this approach to
find the distance distribution of point lenses in the {\it Spitzer} sample,
which acts as the ``denominator'' in determining the planet frequency
as a function of distance.

Nevertheless, while these arguments and methods are quite adequate
for determining the statistical properties of the lens populations, they
obviously can fail catastrophically in individual cases.  The possibilities
of such failures, whether catastrophic or not, is of greater concern
for planetary detections for two reasons.  First, there are many fewer planets
than point lenses,
so information about each one is intrinsically more valuable.  Second,
planets have other measurable parameters, namely their mass ratio $q$
and their projected separation in units of the Einstein radius $s$.
Full interpretation of these other parameters requires a mass and distance
measurement.

Here, we report on the second {\it Spitzer} planet
that lacks a $\theta_\e$ measurement, OGLE-2016-BLG-1067Lb.
This planet joins the very first \emph{Spitzer} microlensing planet, 
OGLE-2014-BLG-0124, which also lacked a $\theta_\e$
measurement, and for which therefore additional techniques 
had to be developed to
constrain the mass and distance \citep{ob140124}.
(A new determination of the mass for this system,
refining the original one in \cite{ob140124}, has been carried out
by \cite{beaulieu17} combining the \emph{Spitzer}-based microlens parallax
with a constraint on the lens flux based upon Keck II 
adaptive optic (AO) observations.) In the case of OGLE-2016-BLG-1067Lb
we show that a mathematical analysis of the light curve alone leads to
an 8-fold degeneracy, in addition to the fact that $\theta_\e$ is not
measured.  Hence, while we draw on the techniques of \cite{ob140124},
we must incorporate other techniques as well, including some that are
ultimately dependent on \cite{hangould95} and \cite{21event}.  In the end,
we are able to identify this as a Saturn-mass planet orbiting a mid M dwarf.


\section{Observations}
\label{sec:obs}

\subsection{Ground observations}

The new microlensing event OGLE-2016-BLG-1067
was first alerted 
by the Optical Gravitational Lensing Experiment
(OGLE) collaboration on June 10 2016, UT~19:32
based on observations
with the OGLE-IV $1.4$~deg$^2$ camera mounted on the $1.3~\mathrm{m}$
Warsaw Telescope at Las Campanas Observatory in Chile
through the Early Warning System (EWS) real-time
event detection software \citep{ews15}. 
The event is located at equatorial coordinates
$\mathrm{R.A.} = 18:12:49.08$, $\mathrm{decl.} = -27:00:45.5$
(corresponding to $(l,b)=(4^\circ.66,-4^\circ.25)$)
in OGLE field BLG523, with a relatively low cadence of 0.5-1
observations per night, mostly in $I$ band, and only
sparse $V$-band data. In this analysis we make use of 
the OGLE re-reduced difference
image analysis photometry \citep{ews03}. 

The microlensing event has also been reported and observed 
by the Microlensing Observations in
Astrophysics (MOA) collaboration with the $1.8~\mathrm{m}$
MOA-II telescope  located at the Mt.~John Observatory in New Zealand
\citep{sumi03}, and named  MOA-2016-BLG-339.
The observations were carried out in the ``MOA-Red'' filter
(a wide $R+I$ filter) with a cadence $\sim 2\,{\mathrm{hr}}^{-1}$;
in addition $V$-band observations have been taken,
in particular  during the decreasing part of the 
microlensing-event magnification.
We will use these data to constrain the color of the source.
The data were reduced using the MOA re-reduced difference
image analysis (DIA) photometry \citep{moa01}. 

Additionally, the event was monitored by the KMTNet lensing survey 
\citep{kmt16} with three identical $1.6~\mathrm{m}$ telescopes 
located at the Cerro Tololo Interamerican
Observatory in Chile (KMTC), South African Astronomical Observatory 
in South Africa (KMTS), and Siding Spring Observatory in Australia (KMTA).
It lies in KMTNet field BLG32, which has a cadence
of $0.4\,{\rm hr}^{-1}$, enabling almost round-the-clock coverage
at reasonably high density.  The KMTNet data, in the $I$ band, are reduced using
the difference imaging algorithm of \cite{albrow09}.

For the OGLE and MOA surveys, we make use of the data
starting from the 2015 season, overall (excluding
a few outliers) 198 and 1463 data points,
respectively; for KMT we use 2016 data (339, 310 and 210 
for KMTC, KMTS and KMTA, respectively), for a total of 2520
ground-based data points.

\subsection{\emph{Spitzer} observations} \label{sec:spz}

The microlensing program with \emph{Spitzer} for 2016, Cycle 12 
of the warm mission \citep{warm10},
was awarded a total of 300 hours \citep{spitzer15a,spitzer15b}.
One part of the project was specifically devoted to the follow up of events
in the K2C9 footprint \citep{k2c9}\footnote{We recall in particular
the analysis of MOA-2016-BLG-290, a single lens low mass star/brown dwarf
in the Galactic bulge with the determination of the  satellite 
microlensing parallax from both K2 and \emph{Spitzer} \citep{zhu17b}.}.
The larger part of the time was allocated with the
aim of determining the Galactic distribution of planets \citep{21event,zhu17a}.
OGLE-2016-BLG-1067 is located outside
of the K2C9 footprint and therefore its selection followed the rules dictated
by the ``Criteria for Sample Selection to Maximize Planet Sensitivity 
and Yield from Space-Based Microlens Parallax Surveys'' \citep{yee15}.
We recall that, on a weekly basis, the list of the events to be followed up
is finalized 4 days prior to the beginning of the observational 
sequence (Figure~1 from \citealt{ob140124}).
\cite{yee15} defined a set of criteria for selecting events,
and the corresponding observing strategy,
according to which they may (or may not) be included
in the sample of events for building up the statistics
for determining the Galactic distribution of planets.
These criteria allow events to be selected ``objectively'' (if they
meet some pre-defined criteria), ``subjectively'' (at the discretion
of the team), or ``secretly''.  ``Objective'' events \emph{must} be
observed by \emph{Spitzer}.  Therefore, planets detected in these events
are included in the Galactic-distribution sample regardless of whether
they  give rise to signatures before or after the time they meet
these criteria.  See for example the analysis by \citet{ob161190} of
OGLE-2016-BLG-1190.  ``Subjective'' events must be publicly
announced, together with a complete specification of the observation
plan.  Hence, planets that give rise to signatures in data that are
available prior to this announcement cannot be included in the sample.
For this reason, it is also possible to choose events ``secretly'', in
case it is unknown whether the event will be promising (and so worth
extended \emph{Spitzer} observations) at the time of the \emph{Spitzer}
upload.  In this case, the event may subsequently be announced as a
``subjective'' event (with public commitment to carry out extended
observations), or dropped.  In the latter case, the planets discovered
in the event  prior to announcement cannot be included in the sample.

OGLE-2016-BLG-1067 fell in the last category.  It was chosen ``secretly''
for the first week of \emph{Spitzer} observations.  Because it lay
far to the East, the \emph{Spitzer} Sun-angle restrictions prevented
it from being observed until near the end of that week, so that only
3 observations were made.  By the decision time for the second week,
it appeared that the event was turning over at low magnification
and so was dropped without making a ``subjective''
announcement.  However, on June 24 UT~16:20, the MOA group announced
an anomaly in this event based on their real-time data analysis.
Based on this, it was decided to resume observations (after a
1-week hiatus), although it was recognized that the planet could
not be included in the sample.

In order to verify the team's original assessment that the planet
could not be included in the Galactic-distribution sample, it is
necessary to determine whether the decision to stop observations
(rather than select the event ``subjectively'') was influenced
by the presence of the planet.  That is, just as planets
cannot be included in the sample if the decision to observe
them is influenced by the presence of the planet, they equally
cannot be excluded from the sample if the decision to stop
observations is made due to its presence.  This is a concern
in the present case because the planet was first recognized from
a {\it dip} in the light curve.  Such a dip could in principle
have been misinterpreted by the team as the event ``turning over''.
Hence, we reviewed the decision process quite carefully.  We
find that the decision was made based on data
${\rm HJD}^\prime \leq 7557.644$, i.e., roughly 4 days before
the onset of the ``dip'' that led to the MOA alert.  The
effect of the planet on the magnification profile during these
earlier observations is far below the observational error bars.
Therefore, we conclude that the presence of the planet did not
in any way influence the team's decision.

Overall we have obtained 25 ``epochs'' of \emph{Spitzer} data,
on average 1 each 24 hours, except during the one-week gap after the first
3 data points. Each epoch is composed by six $30~\mathrm{s}$
dithered exposures. For the observations we
use the $3.6~{\mu}\mathrm{m}$ channel 1 of the IRAC camera
\citep{fazio04}. The data reduction follows the
specific pipeline described in \cite{novati15b}.

\section{Light Curve Analysis} \label{sec:mcmc}

\begin{figure}
\plotone{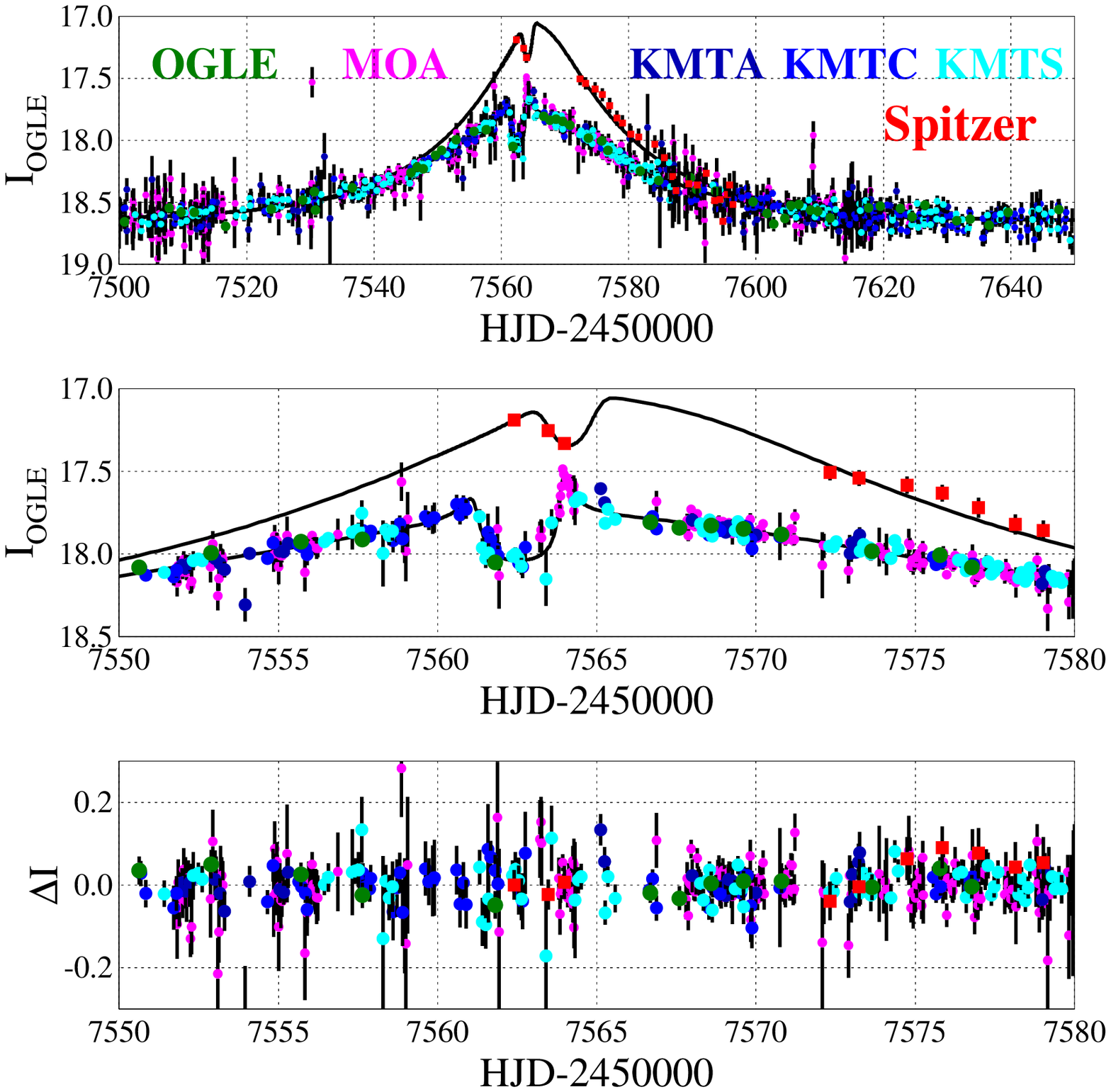}
\caption{Light curve data of OGLE-2016-BLG-1067 (top panel),
zoom around the anomaly (middle panel),
and residual light curve (bottom panel).
\emph{Spitzer} and ground base data have
square and circle symbols, respectively.
The data from the different data sets
are color coded. The model and the residual light curve
are for the $\pi_\mathrm{E,-,-}$-large-$s$ solution,
the best model according to the $\chi^2$ hierarchy
(Table~\ref{tab:large_s}) with $\chi^2$/dof=2562/2545.}.
\label{fig:cl_04}
\end{figure}

\begin{figure}
\plotone{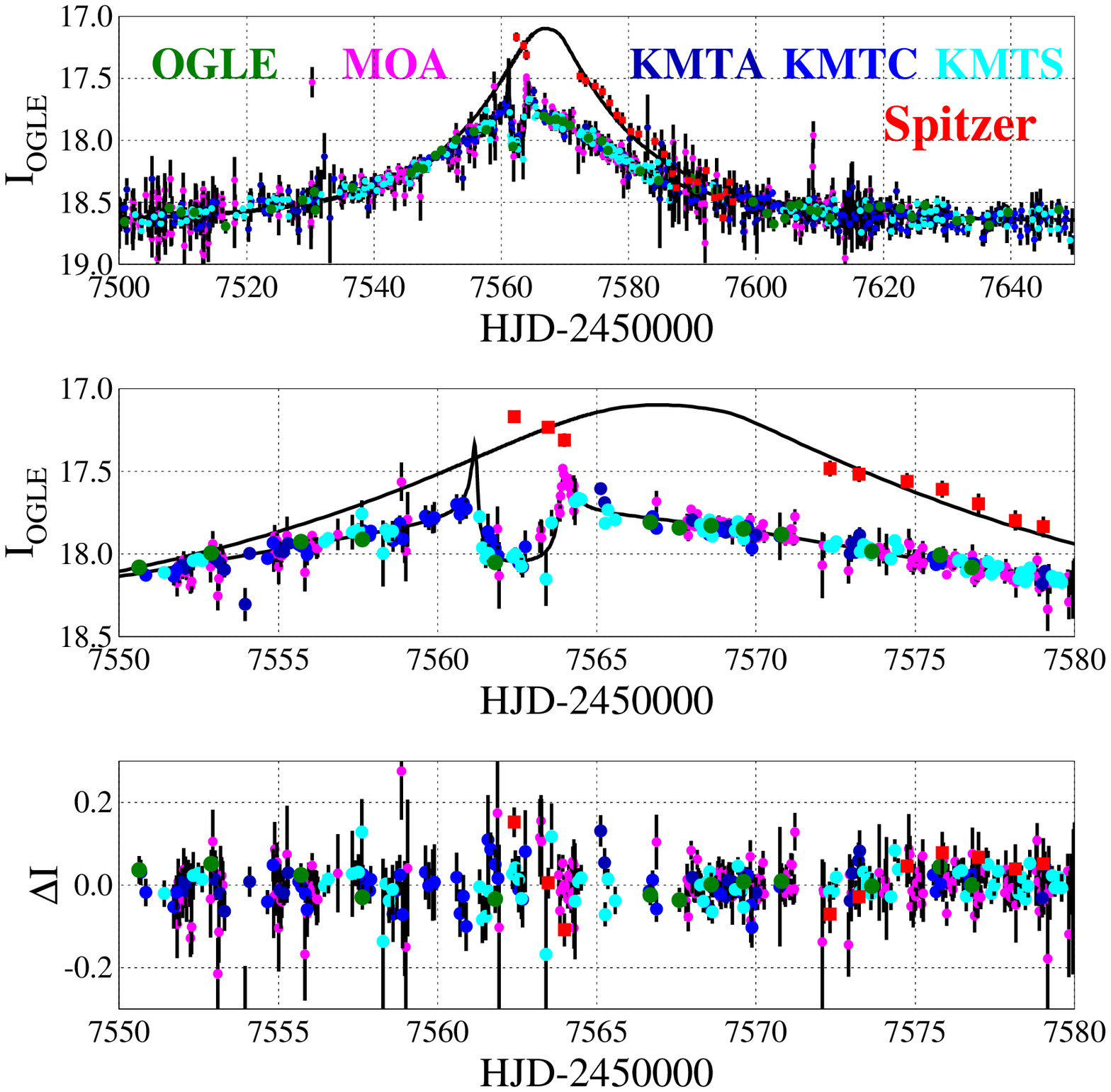}
\caption{Light curve data of OGLE-2016-BLG-1067 (top panel),
zoom around the anomaly (middle panel),
and residual light curve (bottom panel).
\emph{Spitzer} and ground base data have
square and circle symbols, respectively.
The data from the different data sets
are color coded. The model and the residual light curves
are for the $\pi_\mathrm{E,+,-}$-small-$s$ solution,
the worst model according to the $\chi^2$ hierarchy
(Table~\ref{tab:small_s}) with $\chi^2$/dof=2591/2545.}.
\label{fig:cl_03}
\end{figure}

\begin{figure}
\plotone{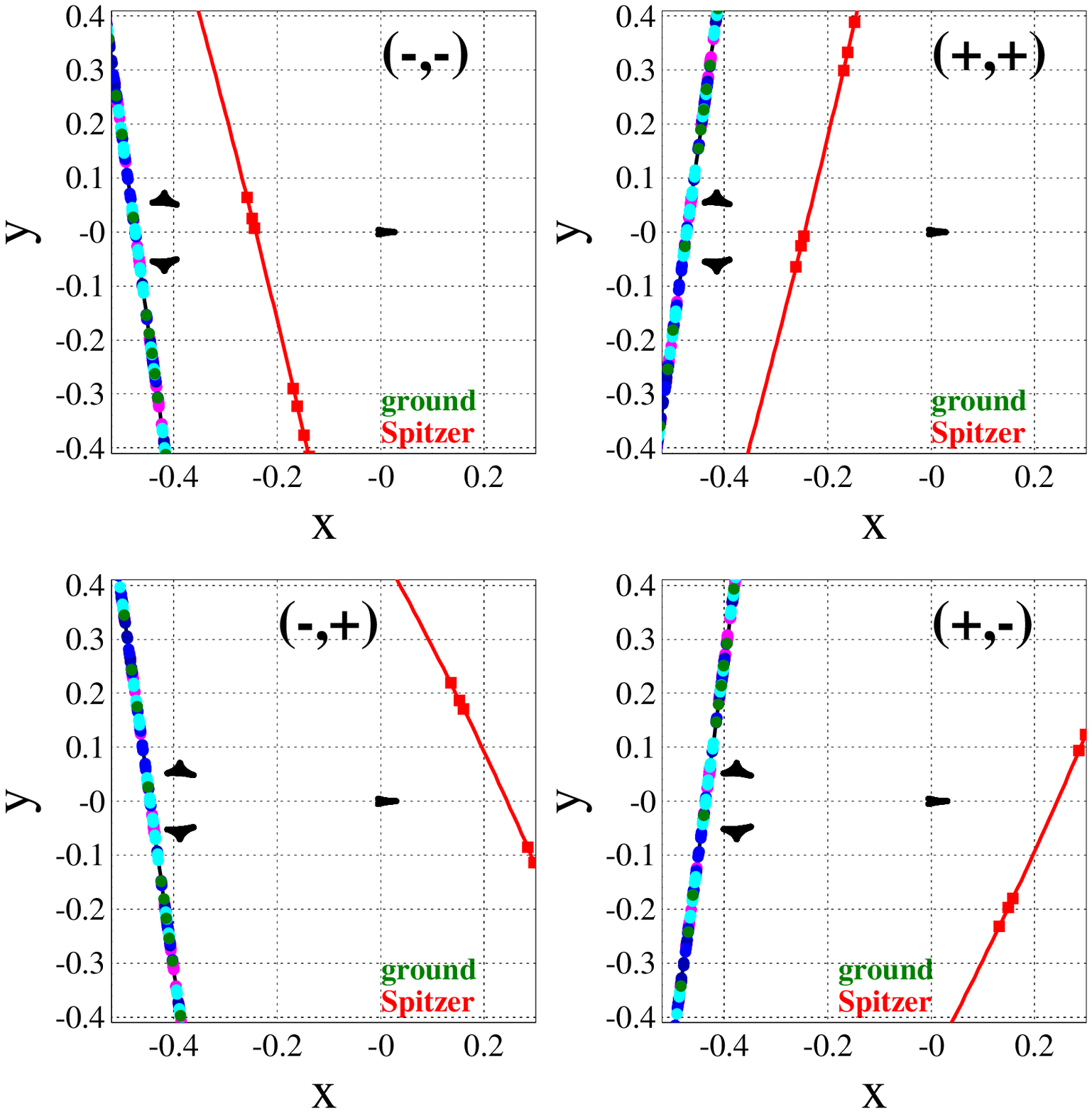}
\caption{The caustic curves and the source
trajectory in the lens plane as seen from ground and from \emph{Spitzer}
for the four large-$s$ solutions (Table~\ref{tab:large_s}).
\emph{Spitzer} and ground base data have
square and circle symbols, respectively.
The data from the different data sets
are color coded as in Figure~\ref{fig:cl_04} and Figure~\ref{fig:cl_03}.
}
\label{fig:caustic_sp}
\end{figure}

\begin{figure}
\plotone{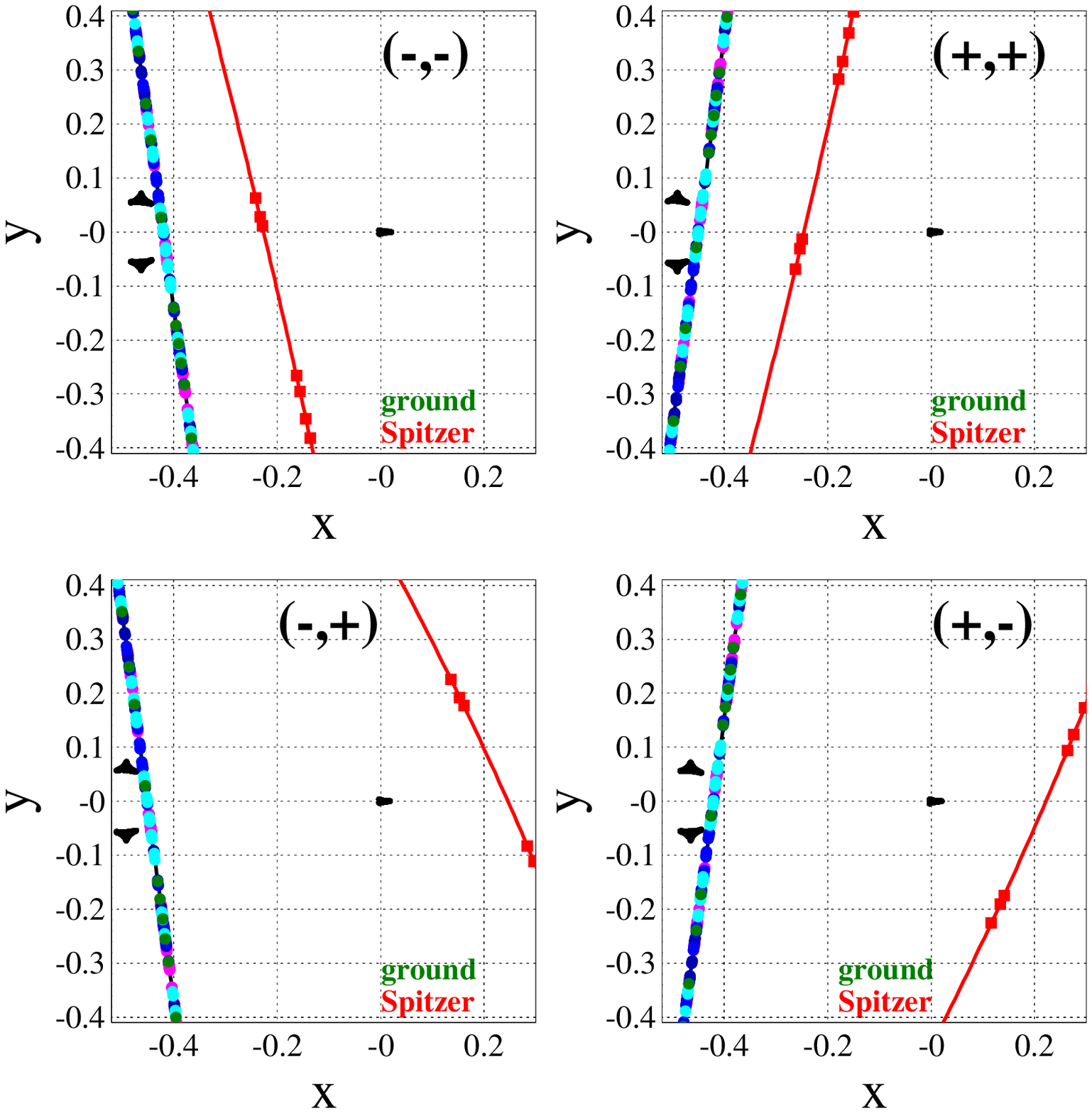}
\caption{The caustic curves and the source
trajectory in the lens plane as seen from ground and from \emph{Spitzer}
for the four small-$s$ solutions (Table~\ref{tab:small_s}).
\emph{Spitzer} and ground base data have
square and circle symbols, respectively.
The data from the different data sets
are color coded as in Figure~\ref{fig:cl_04} and Figure~\ref{fig:cl_03}.
}
\label{fig:caustic_sm}
\end{figure}

The light curve of the event mostly follows a single-lens model except 
for a deviation occurring at about peak magnification.
The single-lens model indicates a low magnification 
event that is punctuated by a short dip (Figures~\ref{fig:cl_04} 
and \ref{fig:cl_03}), 
which is the classic signature of
a ``minor image'' perturbation due to a planet.  The host star gives
rise to two images, which according to Fermat's principle are at
stationary points of the time-delay surface.  The smaller of these two
images is at a saddle point and so can easily be annihilated if a planet
lies at or close to this position.  The ratio of the unperturbed
magnification of these two images is $(A-1)/(A+1)$, where $A$ is
the total magnification.   Hence for a low-magnification $A_{\rm max}\sim 2.5$
event such as this one, at most a fraction $\sim 30\%$ of the flux can
be eliminated.  Moreover, the point where the flux is most strongly
suppressed is flanked by two triangular caustics; however, the light curve
does not exhibit any caustic crossings.  Rather, it shows signs of
cusp approaches just before and after the ``dip''.  Hence, we conclude
that the source has passed close to, but has not intersected the two
caustics that flank the dip in the magnification profile.
This introduces a potential degeneracy with the source trajectory
passing on either side on the planetary caustics
with respect to the central caustic (Figures~\ref{fig:caustic_sp} 
and \ref{fig:caustic_sm}).

The microlensing magnification $A(t)$ for a single lens is,
in the standard \cite{pacz86} form,
a function of three parameters: 
the time of maximum magnification, $t_0$,
the impact parameter, $u_0$, and the Einstein time, $t_\mathrm{E}$.
Additionally, the effect of finite source size is
parameterized by $\rho=\theta_*/\theta_\mathrm{E}$, where $\theta_*$
is the angular source size
and $\theta_\mathrm{E}$ is the Einstein angular radius.
To model a binary lens system we introduce
three additional parameters:
the mass ratio between the planet and its host star,
$q$, their instantaneous projected separation,
in units of the Einstein radius, $s$,
and an angle specifying the source trajectory
with respect to the binary axis, $\alpha$.
In addition to this set of seven non-linear 
parameters, for a given model, there are
two  flux parameters, the source flux, $f_s$,
and the blend, $f_b$, for each data set,
entering linearly in the magnification model,
$f(t) = f_b+ f_s \cdot A(t)$.

From the observed ground-based light curve
we may obtain a first guess
on the values of the binary parameters based on the single 
lens model, for which $u_0\sim 0.4$ and $t_\mathrm{E}\sim 30~\mathrm{d}$,
and the expected planetary model.
Because of the {absence of a caustic crossing} and the anomaly
occurring at about the peak magnification
we expect $u_0=s-1/s$, giving the pair of solutions $s\sim 0.8$
and $s\sim 1.2$. The anomaly shape, a clear dip,
indicates without ambiguity that only the first, close,
solution is viable. For a trajectory approximately perpendicular
to the binary axis, $\Delta t/t_\mathrm{E}\sim 2 \eta_{c,0}$,
where $\eta_{c,0}\simeq 2 \sqrt{q}/s \cdot (1-0.5\,s^2)$ 
is the position of the planetary caustic
along the axis perpendicular to the binary axis \citep{han06},
and $\Delta t$ is the dip duration. For $\Delta t\sim 3~\mathrm{day}$
we evaluate therefore $q\sim 10^{-3}$.
We note that this is the same preferred solution as 
in the Real-Time Microlensing Modeling by 
V.~Bozza\footnote{http://www.fisica.unisa.it/gravitationAstrophysics/RTModel/2016/RTModel.htm.},
which is the result of a completely independent 
and automated search algorithm across the full parameter space.
(Comparing to the real-time models by V.~Bozza,
we note that these did not include KMTNet data.
In the present analysis the dip, and therefore
the solution with $s<1$, is much better constrained
because of the dense coverage ensured by these data.)

In addition to the basic lensing parameters, 
the simultaneous observations from space with \emph{Spitzer} allow us to
constrain the microlensing parallax,
which we parameterize with the two components
along the North and East axes, $\pi_\mathrm{E,N},\,
\pi_\mathrm{E,E}$ \citep{gould04}. For two fixed observers,
the microlensing parallax 
is affected by a four-fold degeneracy \citep{refsdal66,gould94},
which in principle can be removed in the case of binary lens systems
(as in the case of OGLE-2015-BLG-1212, \citealt{bozza16}).
As we detail below however, also because of the gap
in \emph{Spitzer} data, we are unable to conclusively break this 
degeneracy from light-curve modeling alone.
To constrain the microlens parallax with 
the simultaneous fit of ground and space-based data,
we follow \cite{gould04} using in particular
the known position of \emph{Spitzer}
relative to Earth as a function
of time from the Horizons Ephemeris 
System\footnote{http://ssd.jpl.nasa.gov/?horizons.}.

We search for the best model in the parameter space
through $\chi^2$ minimization 
(for \emph{Spitzer} we add a penalty term
related to a constraint on the flux
that we obtain from color-color regression, see below).
To this purpose, as well as for the determination
of the uncertainties on the parameters, 
we make use of a Markov chain Monte Carlo (MCMC),
which we seed according to the heuristic analysis
presented above. For the modeling we proceed in the geocentric
frame (with $t_{0,\mathrm{par}}=7564$, i.e., about
the time of maximum magnification). 
In order to evaluate the microlensing magnification given the model,
we make use of a combination of codes:
the contour integration \citep{gouldgaucherel97} 
as developed by \cite{bozza10} and recently released to the 
public\footnote{http://www.fisica.unisa.it/GravitationAstrophysics/VBBinaryLensing.htm.}, in the anomaly region, and 
hexadecapole, quadrupole 
or monopole approximations \citep{pejcha09b,gould08} elsewhere.
For the finite source size we adopt \citep{yoo04} 
linear limb darkening coefficients
$\Gamma_{R,\mathrm{MOA}}=0.494$ 
(estimated out of the average of $\Gamma_R$ and $\Gamma_I$),
$\Gamma_I=0.410$ and $\Gamma_{L}=0.144$ based
on the source characterization described below
and the model of \cite{claret11}.
However, the lack of caustic crossings 
make the results only weakly dependent 
on these parameters.

As mentioned, the flux parameters (source and blend flux
for each data set) enter the magnification model linearly.
It is therefore common practice to fit for them analytically within the MCMC 
after the non-linear parameters are fixed in each trial.
This increases the speed but comes at the price of losing
the information of possible covariance terms
from their cross-correlation with the non-linear parameters.
As an exception, here we consider the \emph{Spitzer} flux parameters
as chain parameters, which thereby allows us 
a more reliable characterization
of the $(R_\mathrm{MOA}-L)$ color, which is crucial
for comparison to the color-color based constraint.

A preliminary analysis with the MCMC confirms
the initial assessment of the lack of caustic crossings
and in particular, as discussed in Section~\ref{sec:intro},
the impossibility to measure the Einstein angular radius, $\theta_\mathrm{E}$,
based on the light curve analysis of the finite source size effect, $\rho$,
together with the characterization of the source size, $\theta_*$
(Equation~(\ref{eq:mu}) and Equation~(\ref{eq:mlens})).
More specifically, we can only establish an upper limit for $\rho$, which
in turn can be translated, given the parallax measurement, into a 
lower limit on the lens mass and an upper limit on its distance.
We further explore this line of reasoning in the following sections
devoted to the analysis of the physical parameters of the lens system.
(It is worth recalling that with particular geometry configurations,
as for OGLE-2016-BLG-1195Lb \citep{bond17,ob161195},
it is indeed possible to obtain a clear
measurement of the finite source parameter, $\rho$,
even in absence of caustic crossings.)

\begin{deluxetable}{lrrrr}
\tablewidth{0pt}
\tablecaption{All solutions (``large''-$s$)\label{tab:large_s}} 
\tablehead{Parameters & \multicolumn{2}{c}{small $\pi_\e$}& \multicolumn{2}{c}{large $\pi_\e$}}
\startdata
$\chi^2$/dof        & 2561/2545 & 2567/2545 & 2573/2545  & 2589/2545 \\ [5pt]
& $(-,-)$ & $(+,+)$ &  $(-,+)$ & $(+,-)$ \\ [5pt] 
\hline
 $t_0$ [HJD-2457564.] & $0.325_{-0.064}^{+0.065}$ & $0.300_{-0.063}^{+0.065}$ & $0.308_{-0.065}^{+0.065}$ & $0.241_{-0.066}^{+0.065}$\\ [5pt]
 $u_0$ & $-0.470_{-0.029}^{+0.029}$ & $0.474_{-0.029}^{+0.030}$ & $-0.451_{-0.034}^{+0.031}$ & $0.444_{-0.030}^{+0.032}$\\ [5pt]
 $t_\mathrm{E}$ [days] & $26.5_{-1.1}^{+1.2}$ & $26.4_{-1.1}^{+1.2}$ & $27.2_{-1.3}^{+1.3}$ & $27.5_{-1.3}^{+1.3}$\\ [5pt]
 $\rho$ [$10^{-3}$] & $<5.5$ & $<5.8$ & $<5.9$ & $<5.2$\\ [5pt]
 $\pi_\mathrm{E,N}$ & $0.220_{-0.014}^{+0.015}$ & $-0.223_{-0.015}^{+0.014}$ & $0.620_{-0.048}^{+0.052}$ & $-0.635_{-0.051}^{+0.048}$\\ [5pt]
 $\pi_\mathrm{E,E}$ & $-0.054_{-0.008}^{+0.006}$ & $-0.010_{-0.008}^{+0.007}$ & $-0.204_{-0.017}^{+0.015}$ & $-0.085_{-0.010}^{+0.009}$\\ [5pt]
 $\alpha$ [rad] & $1.427_{-0.006}^{+0.006}$ & $4.854_{-0.006}^{+0.006}$ & $1.426_{-0.006}^{+0.006}$ & $4.853_{-0.006}^{+0.007}$\\ [5pt]
 $s$ & $0.812_{-0.012}^{+0.012}$ & $0.811_{-0.012}^{+0.012}$ & $0.820_{-0.014}^{+0.014}$ & $0.824_{-0.014}^{+0.013}$\\ [5pt]
 $q$ [$10^{-3}$] & $1.460_{-0.055}^{+0.063}$ & $1.462_{-0.054}^{+0.065}$ & $1.463_{-0.054}^{+0.063}$ & $1.467_{-0.054}^{+0.062}$\\ [5pt]
\hline
 $f_\mathrm{s,OGLE}$ & $0.516_{-0.045}^{+0.049}$ & $0.522_{-0.046}^{+0.050}$ & $0.486_{-0.047}^{+0.054}$ & $0.476_{-0.045}^{+0.051}$\\ [5pt]
 $f_\mathrm{b,OGLE}$ & $0.030_{-0.048}^{+0.045}$ & $0.024_{-0.050}^{+0.046}$ & $0.059_{-0.054}^{+0.047}$ & $0.069_{-0.050}^{+0.045}$\\ [5pt]
 $R_\mathrm{MOA}-L_{Spitzer}$ & $1.131_{-0.050}^{+0.050}$ & $1.150_{-0.050}^{+0.049}$ & $1.110_{-0.059}^{+0.059}$ & $1.122_{-0.060}^{+0.060}$\\ [5pt]
\hline
 $\theta_*$ [$\mu$as] & $0.851_{-0.038}^{+0.039}$ & $0.856_{-0.038}^{+0.040}$ & $0.826_{-0.041}^{+0.045}$ & $0.817_{-0.040}^{+0.042}$\\ [5pt]
\hline
\enddata
\end{deluxetable}

\begin{deluxetable}{lrrrr}

\tablewidth{0pt}
\tablecaption{All solutions (``small''-$s$)\label{tab:small_s}} 

\tablehead{Parameters & \multicolumn{2}{c}{small $\pi_\e$}& \multicolumn{2}{c}{large $\pi_\e$}}
\startdata
$\chi^2$/dof        & 2566/2545 & 2572/2545 & 2575/2545  & 2590/2545 \\ [5pt]
& $(-,-)$ & $(+,+)$ &  $(-,+)$ & $(+,-)$ \\ [5pt] 
\hline
 $t_0$ [HJD-2457564.] & $0.312_{-0.063}^{+0.063}$ & $0.289_{-0.062}^{+0.062}$ & $0.300_{-0.062}^{+0.064}$ & $0.231_{-0.062}^{+0.063}$\\ [5pt]
 $u_0$ & $-0.442_{-0.031}^{+0.029}$ & $0.442_{-0.029}^{+0.031}$ & $-0.444_{-0.033}^{+0.031}$ & $0.440_{-0.030}^{+0.033}$\\ [5pt]
 $t_\mathrm{E}$ [days] & $27.6_{-1.2}^{+1.3}$ & $27.5_{-1.2}^{+1.3}$ & $27.4_{-1.3}^{+1.3}$ & $27.6_{-1.3}^{+1.3}$\\ [5pt]
 $\rho$ [$10^{-3}$] & $<4.8$ & $<4.7$ & $<4.9$ & $<4.4$\\ [5pt]
 $\pi_\mathrm{E,N}$ & $0.196_{-0.016}^{+0.016}$ & $-0.198_{-0.017}^{+0.015}$ & $0.611_{-0.047}^{+0.050}$ & $-0.629_{-0.052}^{+0.048}$\\ [5pt]
 $\pi_\mathrm{E,E}$ & $-0.054_{-0.009}^{+0.008}$ & $-0.015_{-0.009}^{+0.008}$ & $-0.202_{-0.016}^{+0.015}$ & $-0.085_{-0.010}^{+0.010}$\\ [5pt]
 $\alpha$ [rad] & $1.430_{-0.006}^{+0.006}$ & $4.851_{-0.006}^{+0.006}$ & $1.431_{-0.006}^{+0.006}$ & $4.847_{-0.006}^{+0.006}$\\ [5pt]
 $s$ & $0.785_{-0.011}^{+0.010}$ & $0.785_{-0.011}^{+0.010}$ & $0.784_{-0.012}^{+0.011}$ & $0.786_{-0.012}^{+0.011}$\\ [5pt]
 $q$ [$10^{-3}$] & $1.317_{-0.031}^{+0.033}$ & $1.321_{-0.032}^{+0.032}$ & $1.322_{-0.031}^{+0.032}$ & $1.332_{-0.032}^{+0.034}$\\ [5pt]
\hline
 $f_\mathrm{s,OGLE}$ & $0.473_{-0.043}^{+0.049}$ & $0.474_{-0.044}^{+0.049}$ & $0.478_{-0.046}^{+0.052}$ & $0.471_{-0.045}^{+0.052}$\\ [5pt]
 $f_\mathrm{b,OGLE}$ & $0.072_{-0.049}^{+0.043}$ & $0.071_{-0.049}^{+0.043}$ & $0.067_{-0.051}^{+0.046}$ & $0.074_{-0.052}^{+0.044}$\\ [5pt]
 $R_\mathrm{MOA}-L_{Spitzer}$ & $1.166_{-0.056}^{+0.055}$ & $1.186_{-0.055}^{+0.054}$ & $1.115_{-0.060}^{+0.059}$ & $1.123_{-0.060}^{+0.062}$\\ [5pt]
\hline
 $\theta_*$ [$\mu$as] & $0.815_{-0.038}^{+0.041}$ & $0.816_{-0.038}^{+0.041}$ & $0.819_{-0.041}^{+0.043}$ & $0.813_{-0.040}^{+0.044}$\\ [5pt]
\hline
\enddata
\end{deluxetable}

Overall we find $(2\times 4)=8$ competitive event
geometries, which are the product of the two degeneracies
anticipated above.
The first degeneracy is driven by ground-based data,
with a larger (smaller) value for $s$ 
for the source trajectory passing 
outside (inside) the planetary caustics,
with respect to the central caustic, with
values about $0.81-0.82$ and $0.78-0.79$, respectively.
(More precisely, the degeneracy occurs
in the $s,q$ parameter space, with $q\sim 1.5\times 10^{-3}$ 
and $1.3\times 10^{-3}$
in the two cases, respectively).
In the second model, the first sharp cusp approach
falls in a gap of the data whereas
both cusp approaches are well sampled
for the first model. 
For each of these two possible source trajectories,
as seen from the ground, we then have the four viable
degenerate microlensing parallax solutions
compatible with the data, 
i.e., the two-fold degeneracy for the microlensing
parallax amplitude, $\pi_\mathrm{E,-}\sim 0.2$
and $\pi_\mathrm{E,+}\sim 0.6$
(the $(-,-)$, $(+,+)$ and $(-,+)$, $(+,-)$
solutions respectively).
For the first solution, $\pi_\mathrm{E,-}$, the source 
trajectory as seen from \emph{Spitzer}
passes in between the central caustic and the planetary caustics, and near 
enough to the latter to show a deviation from the smooth single lens shape.
In the second solution, $\pi_\mathrm{E,+}$,
the trajectory passes far away  from all the caustics.
Although the first three data points along the
\emph{Spitzer} light curve before the gap
hint at a non-Paczy{\'n}ski shape, by themselves they are not sufficient
to unambiguously resolve the degeneracy.

In Tables~\ref{tab:large_s} and \ref{tab:small_s}, we report 
for each of the eight solutions
(for the outer and inner ground-based source trajectory, respectively,
to which we will hereafter refer
as ``large'' and ``small''-$s$)
the minimum $\chi^2$, the best (median) value and the corresponding 16\%-84\%
ranges of the non-linear parameters of the magnification
model, the OGLE flux parameters and the 
$(R_\mathrm{MOA}-L)$ color. As discussed,
the model can not constrain $\rho$,
and we report a $3\,\sigma$ upper limit.
In addition, we also report
the MCMC outcome for the angular source radius,
which is a derived quantity based
on the model-independent prior knowledge of the source color
(see Section~\ref{sec:cmd}). The $\chi^2$ difference does not 
break the degeneracy in the $s$ parameter space.
Indeed, for each pair of degenerate microlensing parallax solutions,
$\Delta\chi^2$ is at most $\sim 5$,
although the large-$s$ solution is systematically favored. 
As for the microlensing parallax degeneracy,
the $\pi_\mathrm{E,-}$ solutions have smaller $\chi^2$.
For both $s$ cases, the $\chi^2$ hierarchy is 
$(-,-),\, (+,+),\, (-,+),\, (+,-)$
with $\Delta\chi^2 \sim 6,\, 12, \, 28$
($\Delta\chi^2 \sim 6,\, 9, \, 24$) for the
large (small)-$s$ solution, respectively,
with the hierarchy being driven by the \emph{Spitzer} data
and only the $(+,-)$ solution clearly disfavored.
The features of the different magnification models are
driven by the microlens parallax for \emph{Spitzer}, whereas
for the ground-based observations, they are
driven by the $(s,\,q)$ pair.
The $\pi_\mathrm{E,-}$ solutions have the peak
magnitication occuring earlier and showing,
although smoothed, the same anomaly seen from ground.
The solutions with large $s$ have, from ground,
both cusp approaches very well sampled by the available
data, whereas those with small $s$ have the first,
sharper, cusp approach falling in a gap of the data.
Finally, each pair of solutions  that differ by the
inversion $u_0 \to -u_0$ is qualitatively indistinguishable.
In Figures~\ref{fig:cl_04} and \ref{fig:cl_03}, we show the light curve
for both ground and \emph{Spitzer} data, together with the
corresponding model and the residuals for the $(-,-)$-large-$s$
and the $(+,-)$-small-$s$ solutions. These two
light curves therefore show,
at least qualitatively, the full range 
of possible configurations. At the same time
they are, respectively, the
best and the worst ones according to the $\chi^2$ hierarchy.
The caustic structure and the source trajectories,
as seen from the ground and from \emph{Spitzer},
are shown for all eight solutions in Figures~\ref{fig:caustic_sp} 
and \ref{fig:caustic_sm}.
An additional analysis of Tables~\ref{tab:large_s} and \ref{tab:small_s}
reveals that except for the degenerate parameters
in the corresponding degenerate solutions,
$s,\,q,\,u_0,\,\pi_{E,N},\,\pi_{E,E}$ and $\alpha$,
the MCMC parameters for all the solutions ,
are compatible with one another at 68\% level.
Additionally, the pair $(s,\, q)$ even for the degenerate solutions
are still compatible at the 90\% level.
Apart from the $\rho$ parameter, for each configuration
the model is very well constrained.
The binary lens topology is extremely
well determined, with relative error
about $2\%$ and $4\%$ in $s$ and $q$, respectively,
and the error of the trajectory angle is $<0.5^\circ$.
The relative error in the microlensing parallax is at most
about $8\%$ and is about $5\%$ and $7\%$ for $t_\e$ and $u_0$, respectively.

\section{Characterization of Source and Lens Fluxes} 
\label{sec:fluxes}

In this Section we carry out the photometric analysis
of the source and discuss the limit that we can put
on the lens flux based on the microlens modeling.

\subsection{Color-Magnitude Diagram} \label{sec:cmd}

The light curve microlensing model gives us
the values for the source and the blend flux.
Additionally, independently from the microlensing model,
we can evaluate the source color. By combining
this information with the analysis of the color-magnitude
diagram (CMD), we can characterize the source,
specifically obtain its angular radius,
and, comparing to the blend, obtain
an upper limit for the lens flux.
Finally, cross-matching the optical
and the \emph{Spitzer} CMDs we can evaluate,
again independently from the light curve model,
a color-constraint between the ground-based
and the \emph{Spitzer} flux, which we can then use
within the light curve modeling.
(This is necessary to account for the incomplete
coverage of the underlying primary
microlensing event with \emph{Spitzer} data, \citealt{21event,zhu17a}).

Following \cite{yoo04}, the key in the color analysis,
with the purpose to obtain the source dereddened color and magnitude,
is the study of the offset of the measured 
to the intrinsic centroid of the ``red giant clump''.
For the latter we have $(V-I,I)_\mathrm{cl,0}=(1.06,14.31)$
\citep{bensby13,nataf13}.
For the first, and for the overall source
color analysis, we rely on the MOA and OGLE data.

\begin{figure}
\epsscale{0.9}
\plotone{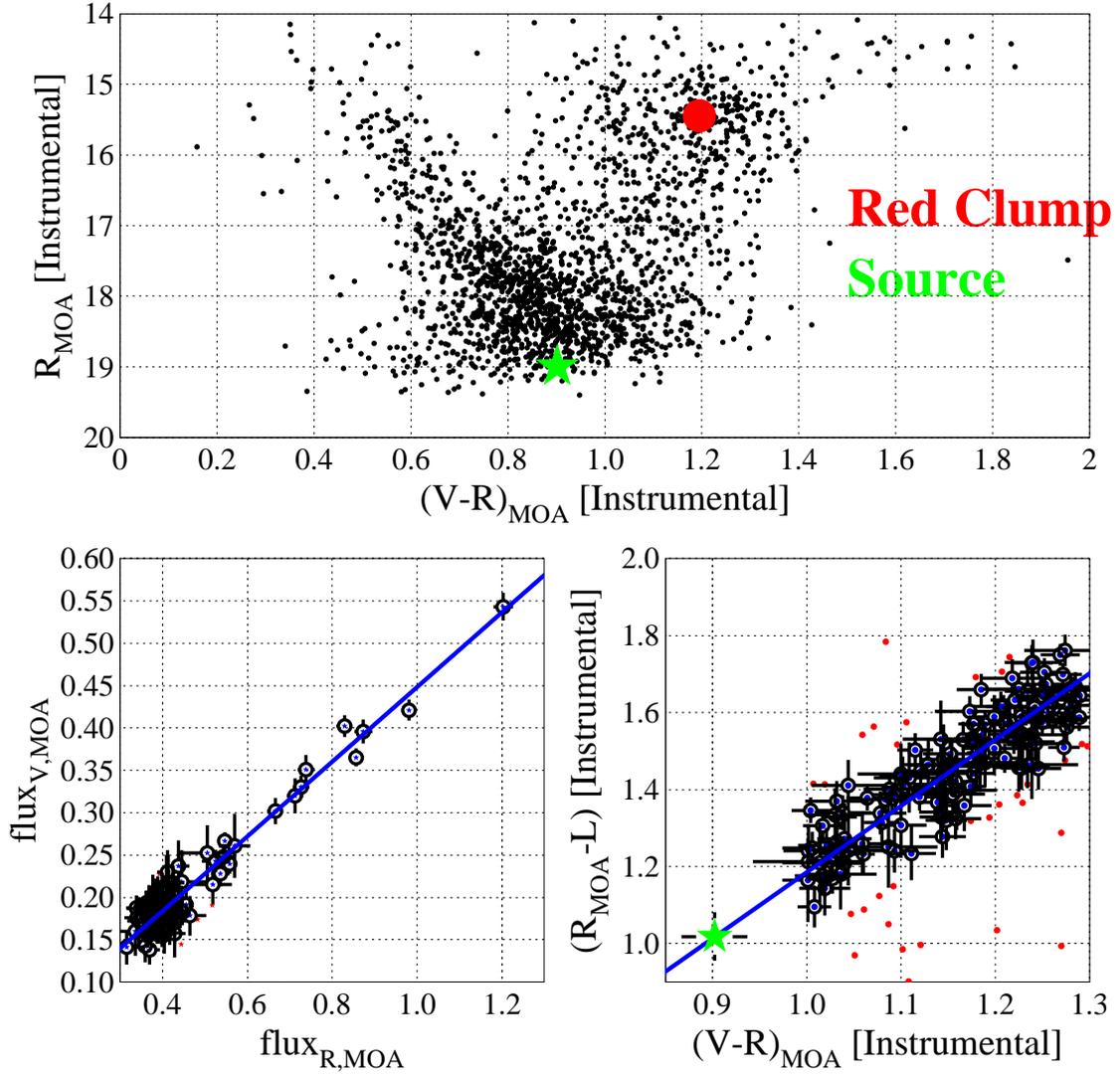}
\caption{Top panel: MOA instrumental CMD of stars 
centered on the event position. The positions of the 
red clump centroid (filled circle) and of the source are indicated.
Bottom left panel: MOA $V$ versus $R_\mathrm{MOA}$ flux for the source color
analysis. The solid line shows the best fit linear model color solution.
Here and in the bottom panel the red dots, without error bars, indicates
the values iteratively rejected within the fit procedure.
Bottom right panel: Instrumental 
MOA and \emph{Spitzer} $(R_\mathrm{MOA}-L)$ vs $(V-R)_\mathrm{MOA}$
and best fit linear model. For the given value of the $(V-R)_\mathrm{MOA}$ source
color, obtained from the data alone (therefore model-independent),
the star indicates the position of the resulting $(R_\mathrm{MOA}-L)$.
}
\label{fig:cmd}
\end{figure}

We start by building the CMD with stars 
centered on the event position based on instrumental 
$V$-band and $R_\mathrm{MOA}$-band magnitudes. 
See Figure~\ref{fig:cmd} (top panel).
In particular following the DIA alignment procedure presented
by \cite{bond17}, we measure the instrumental color
$(V-R)_\mathrm{cl,MOA}=1.195\pm 0.015$,
which we translate, using the MOA calibration 
to the OGLE-III database \citep{ogle3}, to $(V-I)_\mathrm{cl}=1.766\pm 0.016$.
Based on the light curve data, we determine the source color from regression
of MOA-$V$ versus MOA-$R$  flux as the source magnification changes
(Figure~\ref{fig:cmd}, middle panel). It is relevant to recall
that this determination is independent from the light curve modeling.
By correcting for the clump offset we obtain
$(V-I)_\mathrm{s,0} = 0.742\pm 0.035$.

Next, we consider the EWS OGLE-IV CMD
for which we evaluate $I_\mathrm{cl,OGLE-IV}=15.110$.
The resulting source magnitude, as inferred from the microlensing model,
is given in Tables~\ref{tab:large_s} and \ref{tab:small_s}, 
where the OGLE source and blend flux values
have a magnitude zero-point of 18.  Assuming that the source lies
behind the same column of dust as the red clump, we obtain for the
(preferred) $(-,-)$ solutions, $I_{\mathrm{OGLE},0}=17.92\pm 0.10$ and
$I_{\mathrm{OGLE},0}=18.01\pm 0.10$ for the large-s and small-$s$
geometries, respectively.
Overall, based on its position in the CMD, the source
appears to be a G5-G6 dwarf.
As already mentioned in Section~\ref{sec:mcmc},
in the analysis to establish the upper limit on the lens flux
based on the measured blend flux, we also conservatively assume the lens
to be behind the same column of dust as the red clump.

Combining these results, moving from $(V-I,I)$ to $(V-K,K)$
by means of standard color relations \citep{bb88}
and using the relation between color
and surface brightness \citep{kervella04},
we can finally estimate $\theta_*$,
which spans the range of values $0.81-0.86\,\mu\mathrm{as}$
for the different models (via the source flux).

Finally, we can use the source color to constrain
the \emph{Spitzer} instrumental flux
relative to the ground-based one.
Specifically, cross-matching the MOA CMD
to \emph{Spitzer} field stars we establish
a $(VR)_\mathrm{MOA}L_{Spitzer}$ color-color relation. 
See the bottom panel of Figure~\ref{fig:cmd}.
Specifically, given the source instrumental color,
we obtain  $(R_\mathrm{MOA}-L_{Spitzer})=1.018\pm 0.063$
from linear regression of a sample of stars
representative of the bulge population
chosen around the clump position.
As is the case for the source color, 
this determination is also
independent of the light curve model.

\subsection{Limit on the Lens Flux} \label{sec:flens}

As  discussed in Section~\ref{sec:mcmc},
the lack of caustic crossings in the lens geometry
renders impossible the measurement 
of the source size parameter, $\rho$.   
This propagates to the measurement of the Einstein
angular radius $\theta_{\rm E}=\theta_*/\rho$
and eventually to the determination of the lens parameters.
(Rather, we obtained only an upper limit on $\rho$, and thus
a lower limit on $\theta_{\rm E} =\theta_*/\rho$.)
We can, however, combine the measurement
of the microlens parallax $\pi_{\rm E}$ with an upper limit on the lens
flux, to indirectly obtain an upper limit on $\theta_\e$
\citep{ob140124} (see also
the more general analysis of \citealt{yee15b}).  That is, 
if $\pi_{\rm E}$ is known, then increasing $\theta_{\rm E}$ leads to both more
massive ($M=\theta_{\rm E}/\kappa \pi_{\rm E}$) and closer 
($\pi_{\rm rel}=\theta_{\rm E}\pi_{\rm E}$) lenses, whose inferred flux
eventually exceeds the limits on lens flux set by the blended light.
This upper limit on $\theta_{\rm E}$ can also be thought of as a 
lower limit on $\rho=\theta_*/\theta_{\rm E}$.
This procedure is in principle always possible. However,
in order to be effective it must happen, as is the case here,
that the blend be faint enough so as to obtain a meaningful limit.

We are going to exploit this possibility
in the simulation that we carry out to 
include the Galactic model in the determination
of the physical parameters, Section~\ref{sec:galactic_model}.
Given the lens mass and distance,
based on a mass-luminosity
relation \citep{baraffe96}, we can estimate the corresponding
lens magnitude, which we can then  
compare with the (OGLE-based) blend magnitude given 
by the microlensing model\footnote{A caveat here is that the OGLE blend flux
that we estimate is related to the baseline flux offset
used to evaluate the DIA magnitude.
To account for this, for the lens flux limit we 
conservatively take the blend flux plus $2\sigma$,
$\sigma$ being the error on the baseline flux.
As a proxy for the error we take 
the rms reported by OGLE for the baseline magnitude,
$0.061\,\mathrm{mag}$. That is, $\sigma=0.03$ in the $ZP=18$ system}.
Accordingly, we eliminate too close, too bright lenses
corresponding to, for a given source angular size,
increasingly low values of $\rho$.

For reference, the threshold magnitude based
on the blend flux, taking into account
the extinction, is about $I\sim 19.5$,
and for a distance of $1,\,2,\,4$ and $6\,\mathrm{kpc}$
this corresponds to a maximum mass of
$0.25, \, 0.48,\, 0.68$ and $0.80\,\mathrm{M_\odot}$, respectively.

\section{A Sub-Jupiter Mass Planet Beyond the Snow Line} 
\label{sec:res}

With a planet to host mass ratio of about $q\simeq 1.3-1.5\times\,10^{-3}$,
the light curve modeling suggests, most
likely, a sub-Jupiter mass planet.  However, because $\theta_{\rm E}$
is only weakly constrained, and because there are eight different
possible topologies (comprised of two groups with substantially different
values of $\pi_{\rm E}$), we cannot translate these results into an
estimate of physical parameters based on the microlensing light curve alone.

We now seek to resolve and/or tighten all these degeneracies, both
continuous and discrete, by combining four types of
parameter measurements/constraints and three 
arguments.  The parameter measurements/constraints are:
\begin{enumerate}
\item[]{(1) Well-measured microlens parameters 
$(t_{\rm E},\pi_{{\rm E},N},\pi_{{\rm E},E},q,s)$ from the MCMC}
\item[]{(2) Function $\Delta\chi^2(\rho)$ derived from the MCMC}
\item[]{(3) Measurement of $\theta_*$}
\item[]{(4) Flux constraint, as described in Section~\ref{sec:flens}.}
\end{enumerate}
The arguments are discussed in detail in Section~\ref{sec:degen} and are
\begin{enumerate}
\item[]{(1) $\chi^2$  hierarchy}
\item[]{(2) ``Rich argument''}
\item[]{(3) Galactic model}
\end{enumerate}

\subsection{{Resolution of the Degeneracies I: Framework}
\label{sec:degen}}

As discussed in Section~\ref{sec:mcmc}
and tabulated in Tables~\ref{tab:large_s} and 
\ref{tab:small_s}, there
is a $(2\times 4)=8$ fold degeneracy of solutions that need to
at least be considered.  In fact, these degeneracies can be further
subdivided (and re-ordered) as a $(2\times 2 \times 2)=8$ product of
($\pi_{\e,+}$ vs.\ $\pi_{\e,-}$)$\times$($\pi_{\e,N}>0$ vs.\ $\pi_{\e,N}<0$)
$\times$($s_{\rm small}$ vs.\ $s_{\rm large}$).
This ordering reflects both the relative importance of the degeneracies
in terms of physical implications for the system and (happily) the
ease with which they are broken.

To break these degeneracies, we consider three independent pieces
of evidence: 1) $\chi^2$ of the best model for each local minimum,
2) ``Rich argument'', and 3) Bayesian inference based on a Galactic
model.  We specifically evaluate to what extent these separate pieces
of evidence support and/or contradict one another.

We begin by assuming that, in the absence of any other consideration,
each of the eight separate minimum should be considered equally likely
to be the location of the correct solution.

\subsubsection{$\chi^2$ hierarchy}
\label{sec:chi2}

The values of $\chi^2$ for each of the eight minima are given in
Tables~\ref{tab:large_s} and \ref{tab:small_s}.
Since, each model contains the same number of dof,
the nominal relative probability of these models is simply
$\exp(-\Delta\chi^2/2)$.  However, first, some of the models differ
by only $\Delta\chi^2=3$ and even those with fundamentally different
physical implications can differ by only $\Delta\chi^2=6$.  Thus, even
taken at face value, the $\chi^2$ differences do not decisively
distinguish between solutions.  Second, it is well known that microlensing
light curves can have low-level systematics that generate spurious
$\chi^2$ differences at these levels.  Thus, depending on the
specific $\chi^2$ differences between alternate solutions, additional
arguments may be needed to distinguish between them.

\subsubsection{``Rich Argument''}
\label{sec:rich}

The ``Rich argument'' \citep{21event}
states that, other things being equal, small parallax solutions are
preferred over large ones by a factor $(\pi_{\e,+}/\pi_{\e,-})^2$, which, for 
the case we will consider below, yields $(\pi_{\e,+}/\pi_{\e,-})^2\simeq 10$.
The reason is that if the true parallax is small, it will generically
give rise to a large-parallax alternate-degenerate solution, but if
the true parallax is large, it will give rise to a small-parallax
alternate solution with only $(\pi_{\e,-}/\pi_{\e,+})^2$ probability.  Of course,
``all other things'' may not ``be equal''.  For example, in the case
of the very massive planet OGLE-2016-BLG-1190Lb \citep{ob161190},
it was conclusively demonstrated, using two independent supplementary
arguments, that the large parallax solution was correct.  
However, in that case, the balance of evidence favored the
larte-parallax solution even in the absence of supplementary
arguments. First, for OGLE-2016-BLG-1190Lb,
the ``Rich-argument'' preference was much smaller, only 2.7 compared to the
value that we will derive below, 10.
Second, $\chi^2$ actually favored the large parallax solution by
$\Delta\chi^2=13$ (and $\Delta\chi^2$ was not one of the independent
arguments).  

The key point is that, in contrast to the $\Delta\chi^2$ argument, which
could in principle be subject to systematic errors, the ``Rich argument''
is purely statistical in nature, and its resulting probability ratio
must be taken at face value.  

\subsubsection{Galactic Model}
\label{sec:galactic_model}

If the model fitting had resulted in unambiguous measurements of $\bpi_\e$ and
$\rho$, then these would yield $\theta_\e=\theta_*/\rho$
(since we were able to measure $\theta_*$ in Section~\ref{sec:cmd}),
and so also $M=\theta_\e/\kappa\pi_\e$ and $\pi_\rel=\theta_\e\pi_\e$.
Then, because the source distance $D_S\simeq 7.66\,\mathrm{kpc}$ is 
also reasonably well known, there would be no need for a Galactic 
model\footnote{This is the distance at the middle of the bar
according to \cite{nataf13} at $(l,b)=(4^\circ.66,-4^\circ.25)$,
the value we will use throughout the analysis.}.

Unfortunately, $\rho$ is not actually measured (although it is constrained
in the sense that increasingly larger values of $\rho$ yield progressively
worse $\chi^2$), while the measurement of $\bpi_\e$ suffers from the traditional
four-fold degeneracy, including a two-fold ambiguity in its amplitude,
$\pi_\e$.

Nevertheless, the information that we do have
1) precise (albeit four-fold degenerate) measurements of $\bpi_\e$,
2) precise measurement of $t_\e$,
3) constraints (albeit weak) on $\rho$,
and 4) constraints on blended light, together act as powerful constraints on
the Galactic model.

For each of the eight solutions, we begin by extracting from the MCMC
the best fit $a_{0,i}$ and covariance $c_{ij}$ of the three measured 
quantities $a_i=(\tilde v_{\hel,N},\tilde v_{\hel,E},t_\e)$.  Here
\begin{equation}
\tilde{\bdv{v}}_\hel = \tilde{\bdv{v}}_\geo + \bdv{v_{\oplus,\perp}} 
= {\bpi_\e\au\over\pi_\e^2 t_\e} + \bdv{v_{\oplus,\perp}} 
\label{eqn:vtildehel}
\end{equation}
where $\bv_{\oplus,\perp}(N,E)=(0.6,29.3)\,\kms$
is Earth's velocity projected on the plane of the sky.

As we describe below, these measurements (together with the constraints
on $\rho$) already rule out bulge lenses.  We therefore consider
disk lenses drawn according to
a \citet{hangould95} model (except with
rotational velocity $v_\rot=235\,\kms$) and source distance $D_S=7.66\,\kpc$
(and specifically with the distance drawn according to $D_l^2 \cdot \rho(D_l)$,
where $\rho(D_l)$ is the spatial distribution along the
given line of sight).
For each simulated event, we draw a mass randomly from a \cite{kroupa01} mass
function.  We then calculate the resulting $\theta_\e=\sqrt{\kappa M\pi_\rel}$,
$\tilde{\bdv{v}}_\hel = \bmu_\hel\au/\pi_\rel$, $\tilde{\bdv{v}}_\geo$ 
(from Equation~(\ref{eqn:vtildehel})), 
$\mu_\geo = (\tilde v_\geo/\tilde v_\hel)\mu_\hel$, $t_\e=\theta_\e/\mu_\geo$,
and $\rho=\theta_*/\theta_\e$.  Using $D_l$, $M$ and a mass-luminosity
relation from \cite{baraffe96} (and a conservative assumption that the lens
lies behind all the dust) we also calculate the $I$-band flux from the lens.

We then evaluate
\begin{equation}
\chi^2_{\rm gal} = \chi^2(\tilde\bv_\hel,t_\e) + \Delta\chi^2(\rho);\qquad
\chi^2(\tilde\bv_\hel,t_\e) = \sum_{i,j=1}^3 (a-a0)_i b_{ij}(a-a0)_j,
\label{chi2_12}
\end{equation}
where $a_i=(\tilde v_{\hel,N},\tilde v_{\hel,E},t_\e)$ and $b\equiv c^{-1}$.
Each trial in the MCMC gives a value of $\chi^2(\rho)$;
the lower envelope of this distribution gives the minimum $\chi^2$
for a gievn value of $\rho$. Thus, we can construct a function
$\Delta\chi^2(\rho)$ from $\min(\chi^2(\rho) | \rho) - \min(\chi^2)$,
to create a $\chi^2$ penalty that increases as the value of $\rho$.
We count all trials, $N_{\rm trial}$, but tabulate only those
that contribute significantly to the total likelihood ($\chi^2_{\rm gal}<20$),
$N_{\rm tabul}$.  We also exclude trials that fail the flux constraint
(Section~\ref{sec:flens}).  We then calculate a mean
likelihood as the sum of weights evaluated by combining
$\chi^2_{\rm gal}$ and the microlensing rate contribution
\begin{equation}
\langle L\rangle = 
{\sum_{i=1}^{N_{\rm tabul}} w_i \over N_{\rm trial}}\,,\quad 
w_i= \exp(-\chi^2_{{\rm gal},i}/2) \times \theta_{\e,i}\,\mu_{\geo,i}\,. 
\label{eqn:likelihood}
\end{equation}
Of course, these mean likelihoods are very small in all cases. This
simply reflects the fact that 
$a_{0,i}=(\tilde v_{\hel,N},\tilde v_{\hel,E},t_\e)$
is well measured, which immediately ``rules out'' the overwhelming
majority of random trials drawn from the Galactic model.  However, the only
matters of concern to us are 1) what is the {\it relative} likelihood between
different solutions, 2) what are the parameters (and errors) of each solution,
and 3) are the parameters of the most likely (or several most likely)
solutions ``reasonable''?  

\subsubsection{Summary of Three Types of Information}
\label{sec:three_types}

\begin{deluxetable}{lcccccccc}
\tablewidth{0pt}
\tablecaption{Resolution of the Degeneracies: the 8 solutions\label{tab:deg}}
\tabletypesize{\footnotesize} 
\tablehead{
\colhead{Parameter}&
\multicolumn{2}{c}{$(-,-)$}& \multicolumn{2}{c}{$(+,+)$}
& \multicolumn{2}{c}{$(-,+)$}& \multicolumn{2}{c}{$(+,-)$}\\ 
\colhead{}&\colhead{``large''-$s$} & \colhead{``small''-$s$} & 
\colhead{``large''-$s$} & \colhead{``small''-$s$} &
\colhead{``large''-$s$} & \colhead{``small''-$s$} &
\colhead{``large''-$s$} & \colhead{``small''-$s$} 
}
\startdata
\hline
$\chi^2-\chi^2_\mathrm{best}$&0.0&5.2&6.5&11.1&11.7&14.0&27.8&29.5\\ 
$(\pi_\e/\pi_{\e,{\rm smallest}})^2$&$1.30\pm0.27$&$1.04\pm 0.23$&$1.26\pm0.26$&
1.00&$10.8\pm 2.4$&$10.5\pm 2.4$&$10.4\pm 2.3$&$10.2\pm 2.3$\\ 
$\langle L\rangle/\langle L\rangle_\mathrm{best}$&0.59&1.00&$9.5\,10^{-4}$&
$2.1\,10^{-3}$&$6.5\,10^{-2}$&$6.4\,10^{-2}$&$1.9\,10^{-3}$&$2.0\,10^{-3}$\\ 
\hline
\enddata
\end{deluxetable}

Table~\ref{tab:deg} summarizes the results of the three types of information
for the eight solutions.  For each degenerate solution,
following the combined effect of the four-fold microlensing parallax
degeneracy and the $(s,\,q)$ topology degeneracy,
for which    details can be
obtained from Tables~\ref{tab:large_s} and \ref{tab:small_s},
we report: the difference
of $\chi^2$ relative to the best model;
the ``Rich argument'' ratio, i.e,, $(\pi_\e/\pi_{\e,{\rm smallest}})^2$ 
relative to the smallest $\pi_\e$;
and finally, the Galactic model likelihood ratio
$\langle L \rangle/\langle L \rangle_{\rm best}$.

\subsection{{Resolution of the Degeneracies II: Application}
\label{sec:degen2}}

We now discuss how these three types of information discriminate
between the three degeneracies.

\subsubsection{Small vs. Large Microlens Parallax}
\label{sec:bigsmallpie}

The small vs large parallax degeneracy 
is the degeneracy between the first two columns of 
Tables~\ref{tab:large_s} and \ref{tab:small_s}
and the last two columns of these tables.  It is the only one of
the three degeneracies that impacts the interpretation of the
internal nature of the system in a major way.  That is, since
$\pi_{\e,-}/\pi_{\e,+}\sim 1/3$ and since all other parameters are
very similar in these solutions, the inferred mass 
$M=\theta_\e/\kappa \pi_\e$ (or range of allowed masses) will be
three times smaller in the first than the second.  

As shown in Table~\ref{tab:deg}, all three arguments significantly 
favor the small parallax solutions.  First, the best small parallax
solution is favored over the best large parallax solution by $\Delta\chi^2=12$.
Second, of course, the ``Rich argument'' (by definition) favors the 
small parallax solution by a ratio 10.  Third, the Galactic model
likelihood ratio also favors this small parallax solution.  
From Table~\ref{tab:deg},
we see that the best small parallax solution has higher mean likelihood than
the best large-parallax solution by a factor 15.  This is primarily
because the large-parallax 
solutions have more-nearby lenses and hence lower accessible
Galactic volume.  Taken together, the three arguments very strongly
favor the small parallax solutions.

\subsubsection{Positive vs.\ Negative $\pi_{\e,N}$}
\label{sec:posnegpien}

Within this small-parallax $(\pi_{\e,-})$ class of solutions, the
solutions with $\pi_{\e,N}>0$ $(-,-)$ are favored over those
with $\pi_{\e,N}>0$ $(+,+)$ by just $\Delta\chi^2\simeq 6$.
This would not be conclusive, even under the assumption of
purely Gaussian statistics.  However, the Galactic model very
strongly favors the $\pi_{\e,N}>0$ $(-,-)$ solution, by a factor over 400.
It is instructive to track exactly why the Galactic model favors
these solutions, in part because this process allows us to understand
why these solutions are not merely ``better'' but also intrinsically
``reasonable''.

The two classes of solutions have very similar amplitudes 
$\tilde v_{\rm hel}\sim 290\,\kms$ and differ primarily in direction.
Before continuing, we note that this projected velocity corresponds
to a heliocentric proper motion,
\begin{equation}
\mu_\hel = {\tilde v_\hel\pi_\rel\over \au} =
0.9\,{\masyr}{\tilde v_\hel\over 290\,\kms}\,{\pi_\rel\over 0.015\,\mas}.
\label{eqn:vtildemu}
\end{equation}

Since bulge-bulge lensing typically yields $\pi_\rel\sim 0.015\,\mas$,
Equation~(\ref{eqn:vtildemu}) implies that the lens is not likely
to be in the bulge.  Typical proper motions of bulge stars are about
$2.7\,\masyr$ in each direction, so that for bulge-bulge lensing,
the relative motion is $\mu\sim 4\,\masyr$.  Of course, for any particular
event it can in principle be smaller, but the prior probability that it
is smaller than some value scales $p\sim(\mu/4\,\masyr)^3$, which
is quite small in the present case.
Moreover, such low proper motions would require 
$\rho = \theta_*/(\mu t_\e) = 0.012(\mu/0.9\,\masyr)^{-1}$.
However, $\rho>0.006$ is ruled out by the fit at least at the $3\,\sigma$ level.
Thus, we do not explicitly consider bulge lenses.

Next, we rotate the projected velocity to Galactic coordinates and
evaluate it in the frame of the Local Standard of Rest (LSR), by
adding $(12,7)\kms$ in the $(l,b)$ directions.  We then find
$\tilde v_\lsr(l,b) = (235,182)\,\kms$ and 
$\tilde v_\lsr(l,b) = (-233,-154)\,\kms$ for the two solutions.
If all disk stars were on a flat rotation curve of velocity
$v_\rot$, and all bulge stars
were at rest with respect to the center of the Galaxy, we would
expect $\tilde v_\lsr(l,b) = ((\pi_s/\pi_\rel)v_\rot,0)$.  
Adopting, $v_\rot=235\,\kms$,
the offset from the ideal case for $\pi_{\e,N}>0$
can be expressed in terms of proper motion by
\begin{equation}
\Delta\bmu_{\pi_{\e,N}>0} = \biggl({v_\rot\over\au}(\pi_l-2\pi_s),
2.7\,\masyr{\pi_\rel\over 0.07\,\mas}\biggr)\,.
\label{eqn:deltamuplus}
\end{equation}
Adopting $D_S = 7.66\,\kpc$, one can see that the first component
can be accommodated by the $1\,\sigma$ peculiar motion of bulge
sources (i.e., without even considering the peculiar motion of
disk lenses) for $3.6\,\kpc<D_L<4.2\,\kpc$, 
while the second component can be similarly accommodated for $D_L>5\,\kpc$.
Hence, even without allowing for measurement errors and peculiar motion
of the lens, this solution presents only mild tension.
However, the corresponding expression for $\pi_{\e,N}<0$ is:
\begin{equation}
\Delta\bmu_{\pi_{\e,N}<0} = \biggl(-6.5\,{\mas\over\rm yr}\,{7.66\,\kpc\over D_L},
-2.3\,\masyr{\pi_\rel\over 0.07\,\mas}\biggr).
\label{eqn:deltamuminus}
\end{equation}
The first requirement cannot be easily accommodated even for $D_L\simeq D_S$
Hence, the $\pi_{\e,N}>0$ solution is strongly preferred by this argument.

\subsubsection{Large vs. Small $s$}
\label{sec:bigsmalls}

Although the large $s$ solution is favored by $\Delta\chi^2=5$
(and also looks substantially nicer because the data appear to track
the model over the caustic), this is only marginal evidence in
its favor.  While the Galactic likelihood favors  the small $s$
solution, this preference is even weaker than that of the $\chi^2$
discriminant.
Hence, the large/small $s$ degeneracy cannot be resolved.
Fortunately this does not significantly impact the conclusions about
the physical nature of the system.

\subsection{Physical Parameters} \label{sec:finparm}

\begin{deluxetable}{lccc}
\tablewidth{0pt}
\tablecaption{Physical parameter: $(-,-)$ solutions\label{tab:res_mm}}
\tablehead{
\colhead{Parameter}&
\colhead{``large''-$s$} & \colhead{``small''-$s$} & 
\colhead{Adopted}
}
\startdata
\hline
$M_\mathrm{host}$ ($\mathrm{M}_\odot$) &$0.28_{-0.10}^{+0.14}$&$0.31_{-0.13}^{+0.16}$&$0.30_{-0.12}^{+0.15}$\\ [5pt]
$M_\mathrm{planet}$ ($\mathrm{M}_\mathrm{Jup}$)  &$0.43_{-0.16}^{+0.21}$&$0.43_{-0.18}^{+0.22}$&$0.43_{-0.17}^{+0.21}$\\ [5pt]
$D_\mathrm{host}$ ($\kpc$) & $3.68_{-0.64}^{+0.65}$&$3.78_{-0.70}^{+0.68}$&$3.73_{-0.67}^{+0.66}$\\ [5pt]
$a_\perp$ (au) & $1.68_{-0.36}^{+0.37}$&$1.71_{-0.42}^{+0.39}$&$1.70_{-0.39}^{+0.38}$\\ [5pt]
$a_\perp/R_\mathrm{snow\,line}$  &$2.21_{-0.41}^{+0.55}$&$2.02_{-0.37}^{+0.60}$&$2.11_{-0.40}^{+0.58}$\\ [5pt]
$\mu_\hel\,(N)$ (mas/yr)  &$7.7_{-2.0}^{+2.4}$&$7.8_{-2.1}^{+2.4}$&$7.7_{-2.0}^{+2.4}$\\ [5pt]
$\mu_\hel\,(E)$ (mas/yr)  &$-0.4_{-1.0}^{+1.6}$&$-0.4_{-1.3}^{+2.0}$&$-0.4_{-1.1}^{+1.8}$\\ [5pt]
\hline
\enddata
\end{deluxetable}

Table~\ref{tab:res_mm} gives the final adopted parameters, which we derive by 
imposing the Galactic model prior described in 
Section~\ref{sec:galactic_model}. 
In particular, to evaluate the physical
parameters of the planetary system
we combine the MCMC binary-lens caustic
topology parameters, $s$ and $q$, with
the lens and distance from the Galactic
model weighted as in Section~\ref{sec:galactic_model}. 
Following the arguments given
in Sections~\ref{sec:bigsmallpie} and \ref{sec:posnegpien}, we exclude
the six topologies that have $\pi_{{\rm E},+}$ and/or $\pi_{{\rm E},N}>0$.
The remaining two topologies, with $(-,-)$ and either larger or small $s$,
have physical parameters that differ by much less than their errors.
Hence, we simply take the unweighted average of these two solutions, both
for the values and the errors.  In addition to reporting the physical
properties of the system, we also report its heliocentric proper motion
to enable comparison with future observations.

The adopted solution given in Table~\ref{tab:res_mm} has an M-dwarf 
($M_L\sim 0.3~\mathrm{M}_\odot$) host in the Galactic disk
($D_L\sim 4.0~\kpc$), with a Saturn-mass planet 
($M_\mathrm{planet}\sim 0.4~\mathrm{M}_\mathrm{Jup}$) at a projected
distance $a_\perp\sim 1.7~\mathrm{au}$, about twice as far the 
snow-line distance 
(adopting $R_\mathrm{snow\,line}=2.7\,\mathrm{au}\,(M/M_\odot)$).
The error budget, relative error about 40\%,
is dominated by the poorly constrained
finite source size effect, which then led us 
to carry out the Bayesian analysis to derive the physical parameters.

For reference, we report the values for the physical parameters
for the remaining
six topologies, again combining the large and small $s$ solutions.
For the $(+,+)$ solution we find $M_L\sim 0.18~\mathrm{M}_\odot$,
$D_L\sim 4.3~\kpc$, $M_\mathrm{planet}\sim 0.27~\mathrm{M}_\mathrm{Jup}$,
and $a_\perp\sim 1.3~\mathrm{au}$. As expected the solutions
with a larger value of the microlensing parallax
yield a closer and less massive lens host (and planet).
Specifically for the $(-,+)$ solution
$M_L\sim 0.12~\mathrm{M}_\odot$,
$D_L\sim 1.9~\kpc$, $M_\mathrm{planet}\sim 0.17~\mathrm{M}_\mathrm{Jup}$
and $a_\perp\sim 0.91~\mathrm{au}$; for the $(+,-)$ solution
$M_L\sim 0.11~\mathrm{M}_\odot$,
$D_L\sim 1.5~\kpc$, $M_\mathrm{planet}\sim 0.17~\mathrm{M}_\mathrm{Jup}$
and $a_\perp\sim 0.84~\mathrm{au}$.

\subsection{Future Refinement} \label{sec:refinement}

Because the vector parallax $\bpi_\e$ is well-measured, a future
determination of the lens-source relative heliocentric proper
proper motion $\bmu_\hel$ would give a precise measurement of 
the lens mass and lens-source relative parallax from
\begin{equation}
M ={\mu_\hel\,t_{\e,\hel}\over \kappa\pi_\e} = {\pi_\rel\over\kappa \pi_\e^2},
\label{eqn:futurem}
\end{equation}
and
\begin{equation}
\bmu_\hel = {\pi_\rel\over {\rm au}}\,\tilde{\bf v}_\hel ;
\qquad
\pi_\rel = {\mu_\hel\,{\rm au}\over \tilde v_\hel},
\label{eqn:futurepirel}
\end{equation}
where $t_{\e,\hel}\equiv t_\e(\tilde v_\geo/ \tilde v_\hel)$.

The first form of Equation~(\ref{eqn:futurem}) is simpler than the
second in that
it relies on direct observables of the microlensing event
$(t_{\e,\hel},\pi_\e)$ and of future resolved imaging of the lens and
source $(\mu_\hel)$.  Since the errors in $\pi_\e$ and $t_\e$ are each
about 10\% (including the degeneracy between the two surviving solutions),
and since these are roughly anti-correlated, this suggests that 
the mass can ultimately be constrained to $\pm 20\%$, provided
that the proper-motion measurement is more precise than this.
Similarly, the second form of Equation~(\ref{eqn:futurepirel})
gives $\pi_\rel$ directly in terms of a microlensing observable 
$(\tilde v_\hel)$ and an observable from future imaging $(\mu_\hel)$.

Indeed, if the errors for both $\tilde {\bf v}_\hel$ and $\bmu_\hel$
were isotropic (equal and uncorrelated), one can show that
there is no more information available than can be derived from the approach
of the previous paragraph.  In fact, as one can see from 
Tables~\ref{tab:large_s} and \ref{tab:small_s}, the errors have
quite different amplitudes in the two directions.  In addition,
the difference between the two solutions is far greater in
$\pi_{\e,N}$ than $\pi_{\e,E}$.  Thus, in principle, there is 
substantially more information in the first form of 
Equation~(\ref{eqn:futurepirel}) than the second, and the resulting
measurement of $\pi_\rel$ could in principle be input into the
second form of Equation~(\ref{eqn:futurem}) to obtain a more
precise estimate of $M$.  

Unfortunately, in this particular case, the direction of proper motion
(almost due north), implies that there is almost no information
coming from the eastward component, which is the better constrained
component of the $\bpi_\e$ measurement.  Hence, we do not expect
any further improvement from using the slightly more complicated
vector formalism.

The one important application of the vector (as opposed to scalar)
proper motion measurement is that it would decisively rule out
(or possibly confirm one of) the other six solutions.  That
is, of the eight solutions in Tables~\ref{tab:large_s} and \ref{tab:small_s}, 
it is only the two surviving solutions that predict lens-source proper
motions directly in the north direction.  As we have described,
we think that it is extremely unlikely that any of these other six
solutions is correct, but the proper motion measurement would confirm
this.

Of course, by separately imaging the lens and source, one could
also constrain the lens mass from its color and magnitude.

Since the source is relatively faint, $I_s\sim 18.8$, it is plausible
that the lens could be separately resolved with current instrumentation
when they are separated by $\sim 60\,$mas, as was done by 
\citet{batista15} for OGLE-2005-BLG-169.  Based on the heliocentric
proper motion estimates in Table~\ref{tab:res_mm}, this could be done
roughly a decade after the event, i.e., about 2026.  Alternatively,
resolution would also be possible at first light 
of AO cameras on next generation (``30 meter'') telescopes.

\section{Conclusions}

In this paper we have reported the discovery
and characterized OGLE-2016-BLG-1067Lb,
a new exoplanet detected through the microlensing
method toward the Galactic bulge. 
The physical parameters of the system
are strongly constrained thanks
to the measurement of the microlensing
parallax made possible by the simultaneous
observations from the ground
(specifically, the survey data from OGLE,
MOA and KMT) and from \emph{Spitzer},
a satellite orbiting the Sun at more than 1~au from  Earth.
The preferred solution is for a $\sim 0.3~\mathrm{M}_\odot$ host 
in the Galactic disk, orbited by a $0.4~\mathrm{M}_\mathrm{Jup}$
planet with projected separation
at about twice the system snow line.

The detailed analysis of the data
leads to an 8-fold degeneracy
in the microlensing parameter space,
with the usual 4-fold microlensing
parallax degeneracy doubled
by a degeneracy (anticipated by \citealt{gaudi97})
in the caustic topology
$(s,q)$ space, due to an ambiguity 
of the source trajectory with respect
to the planetary caustics of the system.
(This 8-fold degeneracy, however,
reduces to a two-fold degeneracy, driven
by the amplitude of the microlensing
parallax, as far as the physical
parameters of the system are concerned).
In addition, the lack of any caustic crossings
only allows us to determine an upper
limit for the finite size source
microlensing parameter which,
given the microlensing parallax,
translates into a lower limit for
the lens (and planetary) mass.
The light curve analysis
already provides us with additional information
on the maximum lens flux, which we can
then turn into an upper limit
for the lens mass. In order to carry out a more detailed
analysis of the physical parameters of the system,
however, we carry out a Bayesian analysis.
Indeed, together with considerations
based on the $\chi^2$ for the different
solutions and the ``Rich argument'',
this also allows us to break the microlensing
parallax degeneracy. In the end 
we are left with the $(s,q)$ degeneracy 
only, which however has no significant impact on 
our knowledge of the physical parameters.

We have also discussed in some detail,
indeed addressing some new theoretical points
along the lines of the analysis in \cite{gould14},
future mass measurement from the analysis
of the proper motion. Specifically, we show that
AO imaging with next generation instruments 
can definitively distinguish among the four degenerate
microlensing parallax solutions,
and so to decisively rule out (or
possibly confirm one of) the
three solutions excluded in the present analysis.

OGLE-2016-1067Lb is the fifth planet
reported from the ongoing \emph{Spitzer}
microlensing campaign after 
OGLE-2014-0124Lb \citep{ob140124},
OGLE-2015-0966Lb \citep{ob150966},
OGLE-2016-1195Lb \citep{ob161195}
and OGLE-2016-1190Lb \citep{ob161190},
and the fourth located in the Galactic disk.
In compliance with the protocol
explained in \cite{yee15}, however,
this planet does not enter the sample
for the analysis of the Galactic distribution of planets.
Indeed, after the first week, the
observations with \emph{Spitzer}
were stopped and only resumed
with knowledge of an ongoing anomaly. 

At the time of writing 51 exoplanets
have been discovered through the microlensing 
method\footnote{https://exoplanetarchive.ipac.caltech.edu.}.
Compared to other detection methods,
microlensing can more easily probe certain key
parts of the exoplanet parameter space \citep{gaudi12}, 
and specifically exoplanets
orbiting faint stars at large separation.
Within this framework, OGLE-2016-BLG-1067Lb adds to the list of sub-Jupiter
($0.2\lesssim m_\mathrm{p}/\mathrm{M}_\mathrm{Jup}\lesssim 1$) planets
orbiting M-dwarfs beyond the snow line discovered via the microlensing
method.  This population was studied in some detail by \cite{fukui15},
who restricted attention to planetary systems for which the lens
mass was constrained by microlens parallax and/or high-resolution
imaging.  They identified five cold sub-Jupiter planets orbiting M-dwarfs
with such mass contraints.  Subsequently, \citet{ob070349} showed
that OGLE-2007-BLG-349L(AB)c contains a sub-Jupiter planet orbiting
a pair of M dwarfs, based on a combination of a ground-based
parallax measurement and direct imaging with the {\it Hubble Space
Telescope}.  Hence, OGLE-2016-BLG-1067Lb is the seventh such planet.

\acknowledgments
Work by WZ, YKJ, and AG were supported by AST-1516842 from the US NSF.
WZ, IGS, and AG were supported by JPL grant 1500811.
Work by YS was supported by an appointment to the NASA Postdoctoral
Program at the Jet Propulsion Laboratory, California Institute of
Technology, administered by Universities Space Research Association
through a contract with NASA.
Work by CH was supported by the grant (2017R1A4A101517) of
National Research Foundation of Korea.
Work by CR was supported by an appointment to the NASA Postdoctoral Program
at the Goddard Space Flight Center, administered by USRA through a contract
with NASA.
The MOA project is supported by JSPS KAKENHI Grant Number JSPS24253004, 
JSPS26247023, JSPS23340064, JSPS15H00781, and JP16H06287.
The OGLE project has received funding from the National Science Centre,
Poland, grant MAESTRO 2014/14/A/ST9/00121 to AU.
This research has made use of the KMTNet system operated by the Korea
Astronomy and Space Science Institute (KASI) and the data were obtained at
three host sites of CTIO in Chile, SAAO in South Africa, and SSO in
Australia.
This work is based (in part) on observations made with the 
Spitzer Space Telescope, which is operated by the Jet Propulsion Laboratory, 
California Institute of Technology under a contract with NASA. 
Support for this work was provided by NASA through an award issued 
by JPL/Caltech.

\bibliographystyle{apj}
\bibliography{biblio}

\begin{thebibliography}{}

\bibitem[\protect\citeauthoryear{{Albrow} et~al.}{{Albrow}
  et~al.}{2009}]{albrow09}
{Albrow}, M.~D., et~al. 2009, \mnras, 397, 2099

\bibitem[\protect\citeauthoryear{{Baraffe} \& {Chabrier}}{{Baraffe} \&
  {Chabrier}}{1996}]{baraffe96}
{Baraffe}, I.,  \& {Chabrier}, G. 1996, \apjl, 461, L51

\bibitem[\protect\citeauthoryear{{Batista} et~al.}{{Batista}
  et~al.}{2015}]{batista15}
{Batista}, V., {Beaulieu}, J.-P., {Bennett}, D.~P., {Gould}, A., {Marquette},
  J.-B., {Fukui}, A.,  \& {Bhattacharya}, A. 2015, \apj, 808, 170

\bibitem[\protect\citeauthoryear{{Beaulieu} et~al.}{{Beaulieu}
  et~al.}{2017}]{beaulieu17}
{Beaulieu}, J.-P., et~al. 2017, arXiv:1709.00806, \apj submitted

\bibitem[\protect\citeauthoryear{{Bennett} et~al.}{{Bennett}
  et~al.}{2016}]{ob070349}
{Bennett}, D.~P., et~al. 2016, \aj, 152, 125

\bibitem[\protect\citeauthoryear{{Bensby} et~al.}{{Bensby}
  et~al.}{2013}]{bensby13}
{Bensby}, T., et~al. 2013, \aap, 549, A147

\bibitem[\protect\citeauthoryear{{Bessell} \& {Brett}}{{Bessell} \&
  {Brett}}{1988}]{bb88}
{Bessell}, M.~S.,  \& {Brett}, J.~M. 1988, \pasp, 100, 1134

\bibitem[\protect\citeauthoryear{{Bond} et~al.}{{Bond} et~al.}{2001}]{moa01}
{Bond}, I.~A., et~al. 2001, \mnras, 327, 868

\bibitem[\protect\citeauthoryear{{Bond} et~al.}{{Bond} et~al.}{2017}]{bond17}
{Bond}, I.~A., et~al. 2017, \mnras, 469, 2434

\bibitem[\protect\citeauthoryear{{Bozza}}{{Bozza}}{2010}]{bozza10}
{Bozza}, V. 2010, \mnras, 408, 2188

\bibitem[\protect\citeauthoryear{{Bozza} et~al.}{{Bozza}
  et~al.}{2016}]{bozza16}
{Bozza}, V., et~al. 2016, \apj, 820, 79

\bibitem[\protect\citeauthoryear{{Calchi Novati} et~al.}{{Calchi Novati}
  et~al.}{2015a}]{21event}
{Calchi Novati}, S., et~al. 2015a, \apj, 804, 20

\bibitem[\protect\citeauthoryear{{Calchi Novati} et~al.}{{Calchi Novati}
  et~al.}{2015b}]{novati15b}
{Calchi Novati}, S., et~al. 2015b, \apj, 814, 92

\bibitem[\protect\citeauthoryear{{Cassan} \& {Ranc}}{{Cassan} \&
  {Ranc}}{2016}]{cassan16}
{Cassan}, A.,  \& {Ranc}, C. 2016, \mnras, 458, 2074

\bibitem[\protect\citeauthoryear{{Chung} et~al.}{{Chung}
  et~al.}{2017}]{ob151482}
{Chung}, S.-J., et~al. 2017, \apj, 838, 154

\bibitem[\protect\citeauthoryear{{Claret} \& {Bloemen}}{{Claret} \&
  {Bloemen}}{2011}]{claret11}
{Claret}, A.,  \& {Bloemen}, S. 2011, \aap, 529, A75

\bibitem[\protect\citeauthoryear{{Fazio} et~al.}{{Fazio}
  et~al.}{2004}]{fazio04}
{Fazio}, G.~G., et~al. 2004, \apjs, 154, 10

\bibitem[\protect\citeauthoryear{{Fukui} et~al.}{{Fukui}
  et~al.}{2015}]{fukui15}
{Fukui}, A., et~al. 2015, \apj, 809, 74

\bibitem[\protect\citeauthoryear{{Gaudi}}{{Gaudi}}{2012}]{gaudi12}
{Gaudi}, B.~S. 2012, \araa, 50, 411

\bibitem[\protect\citeauthoryear{{Gaudi} \& {Gould}}{{Gaudi} \&
  {Gould}}{1997}]{gaudi97}
{Gaudi}, B.~S.,  \& {Gould}, A. 1997, \apj, 477, 152

\bibitem[\protect\citeauthoryear{{Gould}}{{Gould}}{1994}]{gould94}
{Gould}, A. 1994, \apjl, 421, L71

\bibitem[\protect\citeauthoryear{{Gould}}{{Gould}}{2000}]{gould00b}
{Gould}, A. 2000, \apj, 542, 785

\bibitem[\protect\citeauthoryear{{Gould}}{{Gould}}{2004}]{gould04}
{Gould}, A. 2004, \apj, 606, 319

\bibitem[\protect\citeauthoryear{{Gould}}{{Gould}}{2008}]{gould08}
{Gould}, A. 2008, \apj, 681, 1593

\bibitem[\protect\citeauthoryear{{Gould}}{{Gould}}{2014}]{gould14}
{Gould}, A. 2014, Journal of Korean Astronomical Society, 47, 215

\bibitem[\protect\citeauthoryear{{Gould}, {Carey}, \& {Yee}}{{Gould}
  et~al.}{2013}]{spitzer13}
{Gould}, A., {Carey}, S.,  \& {Yee}, J. 2013, {Spitzer Microlens Planets and
  Parallaxes}, Spitzer Proposal, Spitzer Proposal ID \#10036

\bibitem[\protect\citeauthoryear{{Gould}, {Carey}, \& {Yee}}{{Gould}
  et~al.}{2014}]{spitzer14}
{Gould}, A., {Carey}, S.,  \& {Yee}, J. 2014, {Galactic Distribution of Planets
  from Spitzer Microlens Parallaxes}, Spitzer Proposal, Spitzer Proposal ID
  \#11006

\bibitem[\protect\citeauthoryear{{Gould}, {Carey}, \& {Yee}}{{Gould}
  et~al.}{2016}]{spitzer16}
{Gould}, A., {Carey}, S.,  \& {Yee}, J. 2016, {Galactic Distribution of Planets
  Spitzer Microlens Parallaxes}, Spitzer Proposal, Spitzer Proposal ID \#13005

\bibitem[\protect\citeauthoryear{{Gould} \& {Gaucherel}}{{Gould} \&
  {Gaucherel}}{1997}]{gouldgaucherel97}
{Gould}, A.,  \& {Gaucherel}, C. 1997, \apj, 477, 580

\bibitem[\protect\citeauthoryear{{Gould}, {Yee}, \& {Carey}}{{Gould}
  et~al.}{2015a}]{spitzer15a}
{Gould}, A., {Yee}, J.,  \& {Carey}, S. 2015a, {Degeneracy Breaking for K2
  Microlens Parallaxes}, Spitzer Proposal, Spitzer Proposal ID \#12015

\bibitem[\protect\citeauthoryear{{Gould}, {Yee}, \& {Carey}}{{Gould}
  et~al.}{2015b}]{spitzer15b}
{Gould}, A., {Yee}, J.,  \& {Carey}, S. 2015b, {Galactic Distribution of
  Planets From High-Magnification Microlensing Events}, Spitzer Proposal,
  Spitzer Proposal ID \#12013

\bibitem[\protect\citeauthoryear{{Gould} \& {Yee}}{{Gould} \&
  {Yee}}{2014}]{gouldyee14}
{Gould}, A.,  \& {Yee}, J.~C. 2014, \apj, 784, 64

\bibitem[\protect\citeauthoryear{{Han}}{{Han}}{2006}]{han06}
{Han}, C. 2006, \apj, 638, 1080

\bibitem[\protect\citeauthoryear{{Han} \& {Gould}}{{Han} \&
  {Gould}}{1995}]{hangould95}
{Han}, C.,  \& {Gould}, A. 1995, \apj, 447, 53

\bibitem[\protect\citeauthoryear{{Han} et~al.}{{Han} et~al.}{2017}]{ob161469}
{Han}, C., et~al. 2017, \apj, 843, 59

\bibitem[\protect\citeauthoryear{{Henderson} et~al.}{{Henderson}
  et~al.}{2016}]{k2c9}
{Henderson}, C.~B., et~al. 2016, \pasp, 128, 124401

\bibitem[\protect\citeauthoryear{{Kervella} et~al.}{{Kervella}
  et~al.}{2004}]{kervella04}
{Kervella}, P., {Th{\'e}venin}, F., {Di Folco}, E.,  \& {S{\'e}gransan}, D.
  2004, \aap, 426, 297

\bibitem[\protect\citeauthoryear{{Kim} et~al.}{{Kim} et~al.}{2016}]{kmt16}
{Kim}, S.-L., et~al. 2016, Journal of Korean Astronomical Society, 49, 37

\bibitem[\protect\citeauthoryear{{Kroupa}}{{Kroupa}}{2001}]{kroupa01}
{Kroupa}, P. 2001, \mnras, 322, 231

\bibitem[\protect\citeauthoryear{{Nataf} et~al.}{{Nataf}
  et~al.}{2013}]{nataf13}
{Nataf}, D.~M., et~al. 2013, \apj, 769, 88

\bibitem[\protect\citeauthoryear{Paczy\'{n}ski}{Paczy\'{n}ski}{1986}]{pacz86}
Paczy\'{n}ski, B. 1986, \apj, 304, 1

\bibitem[\protect\citeauthoryear{{Pejcha} \& {Heyrovsk{\'y}}}{{Pejcha} \&
  {Heyrovsk{\'y}}}{2009}]{pejcha09b}
{Pejcha}, O.,  \& {Heyrovsk{\'y}}, D. 2009, \apj, 690, 1772

\bibitem[\protect\citeauthoryear{{Refsdal}}{{Refsdal}}{1966}]{refsdal66}
{Refsdal}, S. 1966, \mnras, 134, 315

\bibitem[\protect\citeauthoryear{{Ryu} et~al.}{{Ryu} et~al.}{2017}]{ob161190}
{Ryu}, Y.-H., et~al. 2017, \apj\, in press, arXiv:1710.09974

\bibitem[\protect\citeauthoryear{{Shvartzvald} et~al.}{{Shvartzvald}
  et~al.}{2015}]{ob151285}
{Shvartzvald}, Y., et~al. 2015, \apj, 814, 111

\bibitem[\protect\citeauthoryear{{Shvartzvald} et~al.}{{Shvartzvald}
  et~al.}{2017}]{ob161195}
{Shvartzvald}, Y., et~al. 2017, \apjl, 840, L3

\bibitem[\protect\citeauthoryear{{Storrie-Lombardi} \&
  {Dodd}}{{Storrie-Lombardi} \& {Dodd}}{2010}]{warm10}
{Storrie-Lombardi}, L.~J.,  \& {Dodd}, S.~R. 2010, in \procspie, Vol. 7737,
  Observatory Operations: Strategies, Processes, and Systems III, 77370L

\bibitem[\protect\citeauthoryear{{Street} et~al.}{{Street}
  et~al.}{2016}]{ob150966}
{Street}, R.~A., et~al. 2016, \apj, 819, 93

\bibitem[\protect\citeauthoryear{{Sumi} et~al.}{{Sumi} et~al.}{2003}]{sumi03}
{Sumi}, T., et~al. 2003, \apj, 591, 204

\bibitem[\protect\citeauthoryear{{Szyma{\'n}ski} et~al.}{{Szyma{\'n}ski}
  et~al.}{2011}]{ogle3}
{Szyma{\'n}ski}, M.~K., {Udalski}, A., {Soszy{\'n}ski}, I., {Kubiak}, M.,
  {Pietrzy{\'n}ski}, G., {Poleski}, R., {Wyrzykowski}, {\L}.,  \& {Ulaczyk}, K.
  2011, \actaa, 61, 83

\bibitem[\protect\citeauthoryear{{Udalski}}{{Udalski}}{2003}]{ews03}
{Udalski}, A. 2003, \actaa, 53, 291

\bibitem[\protect\citeauthoryear{{Udalski}, {Szyma{\'n}ski}, \&
  {Szyma{\'n}ski}}{{Udalski} et~al.}{2015}]{ews15}
{Udalski}, A., {Szyma{\'n}ski}, M.~K.,  \& {Szyma{\'n}ski}, G. 2015, \actaa,
  65, 1

\bibitem[\protect\citeauthoryear{{Udalski} et~al.}{{Udalski}
  et~al.}{2015}]{ob140124}
{Udalski}, A., et~al. 2015, \apj, 799, 237

\bibitem[\protect\citeauthoryear{{Yee}}{{Yee}}{2015}]{yee15b}
{Yee}, J.~C. 2015, \apjl, 814, L11

\bibitem[\protect\citeauthoryear{{Yee} et~al.}{{Yee} et~al.}{2015}]{yee15}
{Yee}, J.~C., et~al. 2015, \apj, 810, 155

\bibitem[\protect\citeauthoryear{{Yoo} et~al.}{{Yoo} et~al.}{2004}]{yoo04}
{Yoo}, J., et~al. 2004, \apj, 603, 139

\bibitem[\protect\citeauthoryear{{Zhu} et~al.}{{Zhu} et~al.}{2016}]{zhu16}
{Zhu}, W., et~al. 2016, \apj, 825, 60

\bibitem[\protect\citeauthoryear{{Zhu} et~al.}{{Zhu} et~al.}{2014}]{zhu14}
{Zhu}, W., {Penny}, M., {Mao}, S., {Gould}, A.,  \& {Gendron}, R. 2014, \apj,
  788, 73

\bibitem[\protect\citeauthoryear{{Zhu} et~al.}{{Zhu} et~al.}{2017a}]{zhu17a}
{Zhu}, W., et~al. 2017a, \aj, 154, 210

\bibitem[\protect\citeauthoryear{{Zhu} et~al.}{{Zhu} et~al.}{2017b}]{zhu17b}
{Zhu}, W., et~al. 2017b, \apjl, 849, L31

\end{thebibliography}

\newpage

\newpage

\newpage

\newpage

\newpage

\end{document}